\documentclass[prd,nofootinbib,preprint,superscriptaddress]{revtex4}
\pdfoutput=1
\usepackage[T1]{fontenc}
\usepackage{amsmath,amssymb}
\usepackage{braket}
\usepackage{epsfig}
\usepackage{graphicx}
\usepackage[usenames,dvipsnames]{color}
\usepackage{subfigure}
\usepackage{slashed}
\usepackage[colorlinks,citecolor=blue]{hyperref}
\usepackage{pdfpages}
\usepackage{color}
\usepackage{comment}
\begin{document}


\title{Filtered cogenesis of PBH dark matter and baryons}

\author{Debasish Borah}
\email{dborah@iitg.ac.in}
\affiliation{Department of Physics, Indian Institute of Technology Guwahati, Assam 781039, India}
\affiliation{Pittsburgh Particle Physics, Astrophysics, and Cosmology Center, Department of Physics and Astronomy, University of Pittsburgh, Pittsburgh, PA 15260, USA}

\author{Indrajit Saha}
\email{s.indrajit@iitg.ac.in}
\affiliation{Department of Physics, Indian Institute of Technology Guwahati, Assam 781039, India}

\begin{abstract}
We propose a novel cogenesis of baryon and dark matter (DM) in the Universe by utilising a first-order phase transition (FOPT) in the dark sector containing an asymmetric Dirac fermion $\chi$. Due to the mass difference of $\chi$ across the bubble walls, it is energetically favourable for $\chi$ to get trapped in the false vacuum leading to the formation of Fermi-ball, which can self-collapse to form primordial black hole (PBH) if $\chi$ has a sufficiently large Yukawa interaction. While such PBH formed out of false vacuum collapse can give rise to the DM in the Universe, a tiny amount of asymmetric $\chi$ leaking into the true vacuum through the bubble walls can transfer the dark asymmetry into the visible sector via decay. The same mass difference of $\chi$ across the two minima which decides the amount of trapping or filtering of $\chi$, also allows $\chi$ decay into visible sector in the true minima while keeping it stable in the false vacuum. Our filtered cogenesis scenario can be probed via FOPT generated stochastic gravitational waves (GW) at near future detectors in addition to the well-known detection aspects of asteroid mass PBH constituting DM in the Universe. 
\end{abstract}
\maketitle

\section{Introduction}
Observations from cosmology and astrophysics not only indicate the presence of dark matter (DM) in our Universe but also the asymmetric nature of visible matter, known as the baryon asymmetry of Universe (BAU) \cite{Zyla:2020zbs, Aghanim:2018eyx}. The fact that the standard model (SM) of particle physics neither has a DM candidate nor can generate the observed BAU has led to several beyond standard model (BSM) proposals in the literature. Among them, the weakly interacting massive particle (WIMP) \cite{Kolb:1990vq, Jungman:1995df, Bertone:2004pz} paradigm  for DM and baryogenesis \cite{Weinberg:1979bt, Kolb:1979qa} as well as leptogenesis \cite{Fukugita:1986hr} for BAU have been the most popular ones. The strikingly similar abundances of DM $(\Omega_{\rm DM})$ and baryon $(\Omega_{\rm B})$ that is, $\Omega_{\rm DM} \approx 5\,\Omega_{\rm B}$ in the present Universe has also led to efforts in finding a common origin or cogenesis of baryon and DM. Some widely studied cogenesis mechanisms include, but not limited to, asymmetric dark matter \cite{Nussinov:1985xr, Davoudiasl:2012uw, Petraki:2013wwa, Zurek:2013wia,DuttaBanik:2020vfr, Barman:2021ost, Cui:2020dly, Borah:2024wos}, baryogenesis from DM annihilation \cite{Yoshimura:1978ex, Barr:1979wb, Baldes:2014gca, Chu:2021qwk, Cui:2011ab, Bernal:2012gv, Bernal:2013bga, Kumar:2013uca, Racker:2014uga, Dasgupta:2016odo, Borah:2018uci, Borah:2019epq, Dasgupta:2019lha, Mahanta:2022gsi}, Affleck-Dine \cite{Affleck:1984fy} cogenesis \cite{Roszkowski:2006kw, Seto:2007ym, Cheung:2011if, vonHarling:2012yn, Borah:2022qln, Borah:2023qag}. Recently, there have also been attempts to generate DM and BAU together via a first-order phase transition (FOPT) by utilising the mass-gain mechanism \cite{Baldes:2021vyz, Azatov:2021irb, Borah:2022cdx, Borah:2023saq, Chun:2023ezg}, phase-separation mechanism \cite{Arakawa:2024bkv} or forbidden decay of DM \cite{Borah:2023god}.

\begin{figure}
    \centering
    \includegraphics[width=0.9\linewidth]{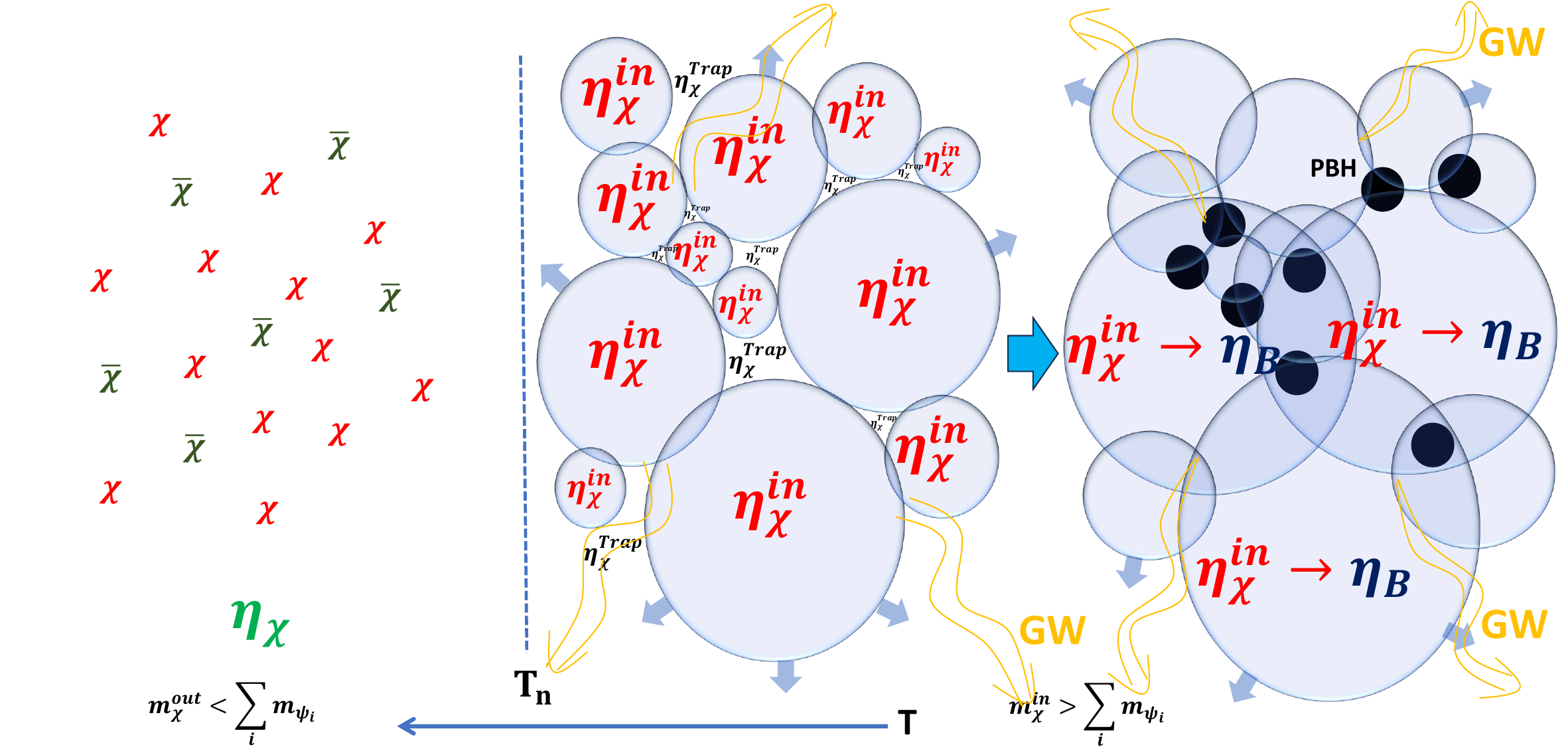}
    \caption{A schematic diagram of the proposed cogenesis scenario.}
    \label{sche}
\end{figure}

While typical baryogenesis or leptogenesis scenarios involve very high scale physics making direct experimental probe difficult, null results at direct detection experiments \cite{LZ:2022lsv} have pushed WIMP DM scenario to a tight corner. This has led to exploration of DM scenarios beyond the WIMP framework. One such promising alternative is primordial black hole (PBH), recent reviews of which can be found in \cite{Carr:2020gox, Carr:2021bzv}. PBH with mass in the range $\sim 10^{17}-10^{22}$ g can be sufficiently stable and free from astrophysical constraints to be a good DM candidate. While PBHs can have their own astrophysical or cosmological signatures, their formation mechanism of also have its own detection aspects. Among well-motivated production mechanisms like PBH from inflationary perturbations \cite{Hawking:1971ei, Carr:1974nx, Wang:2019kaf, Byrnes:2021jka, Braglia:2022phb, Wu:2021gtd}, first-order phase transition \cite{Crawford:1982yz, Hawking:1982ga, Moss:1994iq, Kodama:1982sf, Baker:2021nyl, Kawana:2021tde, Huang:2022him, Hashino:2021qoq, Liu:2021svg, Gouttenoire:2023naa, Lewicki:2023ioy, Gehrman:2023qjn, Kim:2023ixo, Borah:2024lml, Goncalves:2024vkj, Ai:2024cka}, the collapse of topological defects \cite{Hawking:1987bn, Deng:2016vzb} etc., we consider the FOPT as a source while also connecting it to the origin of BAU in the Universe. This leads to a cogenesis mechanism indirectly testable via FOPT generated stochastic gravitational waves (GW). In particular, we consider the collapsing Fermi-ball scenario \cite{Kawana:2021tde, Huang:2022him, Marfatia:2021hcp, Tseng:2022jta, Lu:2022jnp, Marfatia:2022jiz, Xie:2023cwi, Gehrman:2023qjn, Kim:2023ixo, Xie:2024mxr} for PBH formation which also lead to the origin of BAU. A FOPT occurs due to a scalar field which gives a new contribution to a dark fermion mass. The same dark fermion $\chi$ with an initial asymmetry gets trapped in the false vacuum due to the mass gap, leading to the formation of Fermi-balls. These Fermi-balls can further collapse to black holes if there exists an attractive Yukawa force among dark fermions. Since dark fermion maintains equilibrium with the bath prior to the FOPT, its momenta follow an equilibrium distribution allowing some of them to leak into the true vacuum. The enhanced mass of dark fermion in the true vacuum allows its decay into the standard model particles thereby transferring the dark asymmetry into the observed baryon asymmetry. The conceptual visualisation of the idea is summarised in the schematic shown in Fig. \ref{sche}. Depending upon the scale of the FOPT compared to the sphaleron decoupling temperature, BAU can be generated by direct transfer of dark asymmetry into baryons or via the leptogenesis route. By considering the initial dark asymmetry to be a free parameter, we find a small allowed parameter space consistent with the cogenesis of PBH dark matter and BAU. Coincidentally, this allowed region with nucleation temperature around a few tens of MeV to a few GeV remains within the sensitivity of $\mu$ARES.

This paper is organised as follows. In section \ref{sec1}, we discuss the particle physics framework leading to a high scale origin of dark asymmetry and a first-order phase transition at low scale. In section \ref{sec2}, we discuss the details of filtering or the number of dark fermions leaking into the true vacuum. In section \ref{sec3} we discuss the main results related to cogenesis of PBH dark matter and baryon asymmetry of the Universe and finally conclude in section \ref{sec5}.

\section{The framework}
\label{sec1}
 We first formulate the theoretical setup such that a non-zero dark sector asymmetry can be generated at high scale. There are various ways of generating dark sector asymmetries as studied in asymmetric dark matter scenarios \cite{Nussinov:1985xr, Kaplan:2009ag, Davoudiasl:2012uw, DuttaBanik:2020vfr, Barman:2021ost, Cui:2020dly, Falkowski:2011xh, Biswas:2018sib,Narendra:2018vfw,Nagata:2016knk, Arina:2011cu, Arina:2012fb, Arina:2012aj, Narendra:2019cyt,Mahapatra:2023dbr,Borah:2022qln, Borah:2023qag, Borah:2024wos}. Here, for simplicity, we outline one such UV completion based on the Affleck-Dine (AD) mechanism \cite{Affleck:1984fy}.

Consider an Affleck-Dine field $\eta$ singlet under the SM gauge symmetry but having a global $U(1)_B$ charge 2. The Dirac fermion $\chi$ has global charge 1. The relevant terms involving $\eta, \chi$ can be written as
\begin{align}
    -\mathcal{L} \supset   Y_D \overline{\chi^c} \chi \eta^{\dagger} + \epsilon m^2_\eta \eta^2+{\rm h.c.}
    \label{eq:L}
\end{align}
 While the Yukawa interaction in Eq. \eqref{eq:L} preserves $U(1)_B$, the $\epsilon \mu^2 \eta^2$ term of the AD field explicitly breaks $U(1)_B$. Due to this explicit baryon number violating term, the cosmological evolution of the AD field leads to a generation of non-zero baryon number, which gets transferred to the dark fermion $\chi$ when $\eta$ decays into them. 
 
\begin{figure}
    \centering
    \includegraphics[width=0.6\linewidth]{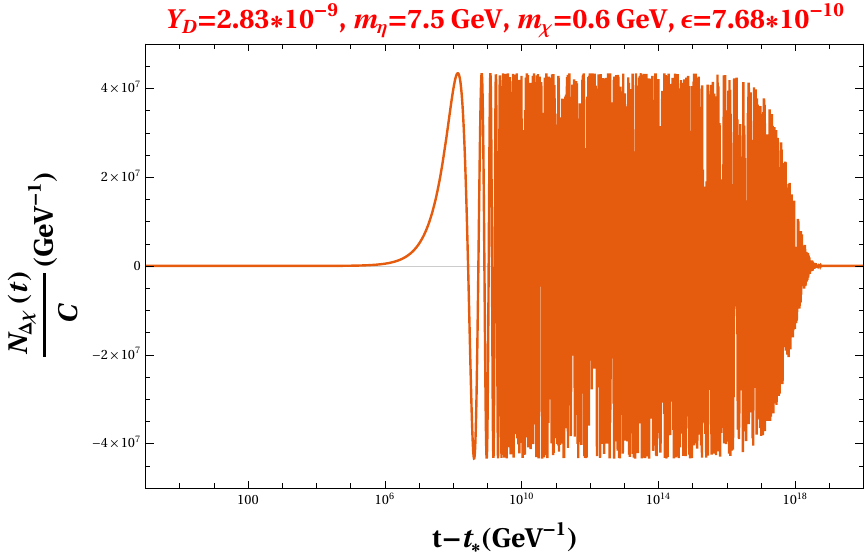}
    \caption{The evolution of the dark sector asymmetry over time is shown for the chosen benchmark values, with the final asymmetry reaching $Y_{\Delta\chi}^{\rm in}\sim 10^{-10}$. }
    \label{fig_asy1}
\end{figure}

\begin{figure}
    \centering
    \includegraphics[width=0.45\linewidth]{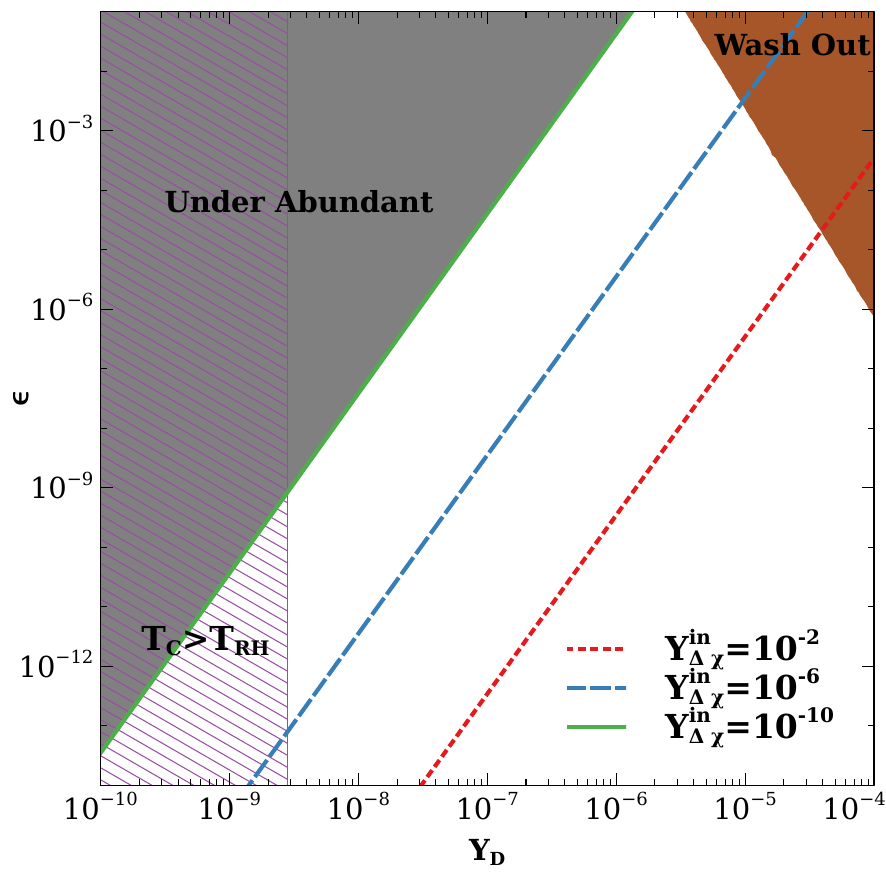}
    \caption{Parameter space in the $Y_D$ vs. $\epsilon$ plane, with contour lines indicating different values of the final dark sector asymmetry for $m_\eta=7.5$ GeV.}
    \label{fig_asy2}
\end{figure}

 Additionally, the $U(1)_B$ preserving AD field potential can be expressed as $V(\eta)= m^2_\eta |\eta|^2  + \lambda_{\eta} |\eta|^4 $. For the field range $\eta^* = \frac{m_\eta}{\sqrt{\lambda_{\eta}}} \lesssim \eta $, the quartic term $\lambda_{\eta} |\eta|^4$ dominates the potential, and the scalar field $\eta$ evolves as $\eta \propto 1/a$. As $\eta$ approaches the value $\eta^*$,  a difference between the real and imaginary components of $\eta$ emerges, leading to an asymmetry in the $\eta$ condensate. The resulting comoving asymmetry generated for $t>t^*$ can then be expressed as \cite{Mohapatra:2021aig, Borah:2022qln} 
\begin{align}
N_{\Delta \chi}(t) &\simeq 4 Q_\eta  A  \eta_{1I}  \eta_{2I} \left( \frac{\eta_I}{\eta^*} \right)
\int_{t^*}^t dt'  \cos[m_1(t' - t^*)]  \cos[m_2(t' - t^*)] e^{-\Gamma_\eta (t' - t^*)}.
\end{align}
Here, $\Gamma_\eta$ denotes the total decay rate of the scalar field $\eta$ into $\chi$ particles. The quantities $\eta_{1I}$ and $\eta_{2I}$ represent the initial values of the real and imaginary components of the field $\eta$, and $\eta_I = \sqrt{(\eta_{1I})^2 + (\eta_{2I})^2}$. The mass parameters are given by $m_1^2 = m^2_\eta - 2A$ and $m_2^2 = m^2_\eta + 2A$, with $A = \epsilon m^2_\eta$.
The asymmetry generated in this process is subsequently transferred to the dark sector via the decay channel $\eta \rightarrow \chi \chi$. Under the assumption $2A \gg \Gamma_\eta m_\eta$, the integral simplifies for $t \gtrsim 1/\Gamma_\eta$ to:
\begin{align}
N_{\Delta \chi}(t) &\simeq C  \frac{\gamma}{8 \epsilon^2 m_\eta},
\label{eqn:Ndensity}
\end{align}
where $\gamma = \Gamma_\eta / m_\eta$ and $C = 4 Q_\eta  A  \eta_{1I}\eta_{2I} \left( \frac{\eta_I}{\eta^*} \right)$. In Fig. \ref{fig_asy1}, we show the evolution of the comoving asymmetry $N_{\Delta \chi}(t)$, which initially grows from zero, oscillates due to mass splitting, and eventually settles to the constant value given by Eq. \eqref{eqn:Ndensity} as the amplitude is exponentially damped for $t \gtrsim 1/\Gamma_\eta$.
 Assuming the decay of the AD field also to reheat the Universe to a temperature $T_{\rm RH}\simeq\sqrt{\Gamma_{\eta}M_{\rm Pl}}$, the dark fermion asymmetry can be estimated as \cite{Lloyd-Stubbs:2020sed,Mohapatra:2021aig, Borah:2022qln}
\begin{equation}
Y^{\rm in}_{\Delta \chi} = \frac{(n_{\chi}-n_{\bar{\chi}})}{s}\simeq \frac{T_{\rm RH}^{3}}{\epsilon m^{2}_\eta M_{\rm Pl}}\label{eq:vis}
\end{equation}
where $m_\eta$ is the mass of the AD field. The dominant washout process is the one which breaks $U(1)_B$ by 4 units: $\chi \chi \leftrightarrow \overline{\chi}~\overline{\chi}$ mediated by $\eta$. Keeping this washout out-of-equilibrium leads to the following condition 
\begin{equation}
T_{\rm RH}^{3}\frac{Y_{D}^{4}\epsilon^2T_{\rm RH}^2}{64 \pi m^{4}_\eta}\lesssim \sqrt{\frac{\pi^2}{90}g_*} \frac{T_{\rm RH}^{2}}{M_{\rm Pl}},\label{eq:washout}
\end{equation}
which can be ensured by suitable choices of couplings and masses. Fig. \ref{fig_asy2} shows the parameter space in $\epsilon-Y_D$ plane consistent with different dark sector asymmetries represented by contours. The grey shaded region is disfavoured as the dark sector asymmetry is less than the observed baryon asymmetry. The magenta meshed region on the left hand side is disfavoured as reheat temperature is less than the scale of FOPT $T_{\rm RH} < T_c$.

 For the first-order phase transition and transfer of dark fermion asymmetry to visible sector in the true vacuum, we consider a minimal framework consisting of a Dirac singlet fermion $\chi$, two singlet real scalars $\Phi_{1,2}$ as the new BSM degrees of freedom present at the scale of FOPT. Another heavy scalar $\zeta$ is introduced in order to allow the coupling of $\chi$ with a fermion $\psi$ in the SM. The low energy effective Yukawa Lagrangian relevant for our study is 
\begin{equation}
    -\mathcal{L} \supset \big ( y_1 \overline{\chi} \Phi_1 \chi + y_2 \overline{\psi} \zeta \chi + {\rm h.c.} \big )+m_\chi \overline{\chi} \chi.
    \label{eq:yukawa}
\end{equation}
While the first term in the above equation leads to a new contribution to dark fermion's mass in true vacuum, the second term paves the way of transferring dark sector asymmetry into SM. As discuss later, the Yukawa coupling $y_2$ and scalar $\zeta$ remains much heavier compared to the FOPT scale and hence do not play any role in low energy phenomenology. The quantum numbers of $\zeta, \psi$ and their roles in cogenesis will be discussed in section \ref{sec3}.

While the singlet scalar $\Phi_1$ drives a first-order phase transition acquiring a non-zero vacuum expectation value (VEV), the other singlet scalar $\Phi_2$ assists in generating the required effective potential while keeping all dimensionless couplings within perturbative limits. We, however, keep the coupling of the SM-like Higgs doublet $H$ with $\Phi_1$ negligible such that the required electroweak symmetry breaking can be achieved independently of this singlet-driven FOPT. With these assumptions, the relevant part of the tree level scalar potential is given by
\begin{align}
   V_{\rm tree} \supset \lambda_1 \left (\Phi^2_1-\frac{v_D^2}{2} \right)^2+ \frac{\lambda_{\Phi_1\Phi_2}}{2}\Phi^2_1 \Phi^2_2 .
    \label{Vtree}
\end{align}
The other possible terms allowed by the symmetries are ignored for simplicity. The dark fermion receives a new contribution to its mass in the true vacuum $\langle \Phi_1 \rangle = v_D$. This splits the zero-temperature mass of $\chi$ inside and outside the bubbles of true vacuum as
\begin{equation}
    m^{\rm in}_\chi = m_\chi + y_1 \frac{v_D}{\sqrt{2}}, \,\, m^{\rm out}_\chi = m_\chi.
\end{equation}
The one-loop correction to the tree-level potential is calculated by considering the couplings of all particles to the singlet scalar $\Phi_1=\phi/\sqrt{2}$. This correction, known as the Coleman-Weinberg (CW) potential \cite{Coleman:1973jx}, can be expressed as
\begin{equation}
V_{\rm CW}(\phi)=\frac{1}{64 \pi^2}\sum_{i=\Phi_2,\chi}n_i m_i^4(\phi)\left \{\log{ \left ( \frac{m_i^2(\phi)}{v_D^2} \right )}-C_i\right \}
\end{equation}
where, $n_{\Phi_2}=1$, $n_\chi=4$ are the respective degrees of freedom
and $C_{\Phi_2,\chi,\zeta}=\frac{3}{2}$. The field dependent masses are
\begin{align*}
m_{\Phi_2}^2(\phi)=\mu_{\Phi_2}^2+\frac{\lambda_{\Phi_1\Phi_2}\phi^2}{2}, \hspace{0.5 cm}
m_\chi^2(\phi)=(m_\chi+y_1\phi/\sqrt{2})^2.  
\end{align*}
Now, the thermal contributions to the effective potential \cite{Dolan:1973qd,Quiros:1999jp} can be written as
\begin{equation}
V_T(\phi,T)=\sum_{i=\Phi_2,\zeta}\frac{n_iT^4}{2\pi^2}J_B \left(\frac{m_i^2(\phi)}{T^2}\right) - \frac{n_\chi T^4}{2\pi^2}J_F \left(\frac{m_\chi^2(\phi)}{T^2}\right)
\end{equation}
where,
$$ J_F(x)=\int_{0}^{\infty}dy\, y^2 \log[1+e^{-\sqrt{y^2+x^2}}], \, J_B(x)=\int_{0}^{\infty}dy\, y^2 \log[1-e^{-\sqrt{y^2+x^2}}].$$
In addition, the Daisy corrections \cite{Fendley:1987ef,Parwani:1991gq,Arnold:1992rz} need to be included in the thermal contribution to enhance the perturbative expansion following the Arnold-Espinosa method \cite{Arnold:1992rz}. These Daisy contributions can be expressed as
\begin{equation}
V_{\rm daisy}(\phi,T)=-\frac{T}{2\pi^2}\sum_{i=\Phi_2} n_i[m_i^3(\phi,T)-m_i^3(\phi)]
\end{equation}
where, $m_i^2(\phi,T)$=$m_i^2(\phi)$ + $\Pi_i(T)$ and the relevant thermal masses are
\begin{align*}
    m_{\Phi_2}^2(\phi,T)= m_{\Phi_2}^2(\phi)+\left(\frac{\lambda_{2}}{4}+\frac{\lambda_{\Phi_1\Phi_2}}{12}\right)T^2.
\end{align*}
The complete effective finite-temperature potential is
\begin{equation}
    V_{\rm eff}(\phi,T)=V_{\rm tree}(\phi) + V_{\rm CW}(\phi) + V_T(\phi,T) + V_{\rm daisy}(\phi,T) .\label{eq:Veff}
\end{equation}
At high temperatures, the Universe stays in the false vacuum $\langle\Phi_1\rangle =0$. At the critical temperature $T_c$ another degenerate minimum appears at $\langle\Phi_1\rangle \neq 0$ with a barrier between them. Subsequently, at a lower temperature known as the nucleation temperature $T_n$, the Universe tunnels through the barrier to go from the false vacuum to the true vacuum $\langle\Phi_1\rangle \neq 0$. The tunneling rate per unit volume defined in terms of O(3) symmetric bounce action $S_3(T)$ is $\Gamma(T)\sim T^4e^{-S_3(T)/T}$\cite{Linde:1980tt}. The nucleation temperature $T_n$ is determined by comparing the tunneling rate with the Hubble expansion rate as
\begin{equation}
    \Gamma(T_n)=\mathcal{H}^4(T_n) = \mathcal{H}_*^4.
\end{equation}
The volume fraction of false vacuum of the Universe is defined by $p(T) = e^{-\mathcal{I}(T)}$ \cite{Ellis:2018mja, Ellis:2020nnr}, where
\begin{align}
    \mathcal{I}(T) = \frac{4\pi}{3}\int^{T_c}_T \frac{dT'}{T'^4}\frac{\Gamma(T')}{{\mathcal H}(T')}\left(\int^{T'}_T \frac{d\tilde{T}}{{\mathcal H}(\tilde{T})}\right)^3.
\end{align}
The percolation temperature $T_p$ is then calculated by using  $\mathcal{I}(T_p) = 0.34$ \cite{Ellis:2018mja} indicating that at least $34\%$ of the comoving volume is occupied by the true vacuum.

The inverse of the FOPT duration is given by $\beta/{\mathcal H}=T\frac{d}{dT} (\frac{S_3}{T})$ while $\alpha_*=\frac{\epsilon}{\rho_{\rm rad}}$ is amount of latent heat released at nucleation temperature where, $\epsilon$ =$\left(  \Delta V_{\rm eff} -\frac{T}{4} \frac{\partial \Delta V_{\rm eff} }{\partial T} \right)_{T=T_n}$, $\Delta V_{\rm eff}= V_{\rm eff}(\phi_{\rm false},T)-V_{\rm eff}(\phi_{\rm true},T)$ and $\rho_{\rm rad}=g_*\pi^2T^4/30$ \cite{Caprini:2019egz}. The bubble wall velocity $v_w$, in general, is related to the Jouguet velocity $v_J = \frac{1/\sqrt{3} + \sqrt{\alpha_*^2 + 2\alpha_*/3}}{1+\alpha_*}$. For the type of FOPT considered here, we can assume $v_w \approx v_J$ \cite{Kamionkowski:1993fg, Steinhardt:1981ct, Espinosa:2010hh}. The calculation for the bounce action $S_3 (T)$ is performed numerically by fitting the effective potential to a generic potential \cite{Adams:1993zs}, the details of which is described in \cite{Borah:2022cdx}.

We assume the dark fermion to have an initial asymmetry with subsequent annihilation of the symmetric part due to $\chi-\Phi_1$ interactions. In order to trap most of the dark fermions in the false vacuum, we impose the condition $T_n \ll m^{\rm in}_\chi-m^{\rm out}_\chi$, such that $\chi$ moving to the true vacuum is kinematically disfavored. However, as the momenta of $\chi$ follows a distribution, some of them can penetrate to the true vacuum. 
We discuss the trapping of $\chi$ in false vacuum and cogenesis in the next two sections.

\section{Filtering of dark fermions}
\label{sec2}
During a first-order phase transition (FOPT), bubbles of the true vacuum nucleate and expand. Within the bubbles, the symmetry is broken, whereas it remains unbroken outside. Consequently, this results in the mass of the particle $\chi$ taking different values inside and outside the bubbles, denoted as $m_{\chi}^{\rm in}$ and $m_{\chi}^{\rm out}$, respectively. As the bubble expands, $\chi$ particles penetrate the bubble wall depending on their energy. Due to the restriction $T_n \ll m^{\rm in}_\chi-m^{\rm out}_\chi$, only the high momentum modes of $\chi$ can penetrate the walls resulting its filtering. The idea of bubble filtering has been used in the context of dark matter and baryogenesis in several earlier works \cite{Baker:2019ndr, Chway:2019kft, Marfatia:2020bcs, Chao:2020adk, Ahmadvand:2021vxs, Baker:2021zsf, Gehrman:2023qjn, Jiang:2023nkj}.

Let $\tilde{v}$ denote the velocity of $\chi$ particles relative to the bubble wall along z direction, while $v_w$ represents the velocity of the bubble wall in the plasma rest frame. The number of particles per unit area going inside the bubble in time $\Delta t$ along the $z$ direction can be written in the bubble wall frame following energy conservation as \cite{Chway:2019kft,Gehrman:2023qjn}
\begin{equation}
    \frac{\Delta N_{\rm in}}{\Delta A}=n_\chi \int \frac{d^3p}{(2\pi)^3}\int_{r_o}^{r_o-\frac{p_z \Delta t}{|p|}}dr \, \mathcal{I}(p) \, \Theta (-p_z) \,f(p,x)
\end{equation}
where $\Delta A$ is the area and the energy conservation is ensured by $\mathcal{I}(p)=\Theta(-p_z-m_d)$, $m_d=m_\chi^{\rm in}-m_\chi^{\rm out}$, $n_\chi$ denotes the degrees of freedom of the particle. Near the bubble wall, $\chi$ particles will follow equilibrium distribution locally and hence $f(p,x)\simeq f_{\rm eq}(p,\tilde{v},T)$. So, the distribution function outside the bubble is
\begin{align}
    f(p,x)&=\frac{1}{e^{\tilde{\gamma}(E-\tilde{v}p_z)/T}+1} \nonumber \\
    & = \frac{1}{e^{\tilde{\gamma}(\sqrt{p^2+(m_\chi^{\rm out})^2}-\tilde{v}p \, \cos{\theta})/T}+1} 
\end{align}
with $\tilde{\gamma}$ being the Lorentz factor. After performing the integration over $r$, the flux of particles entering the bubble $J_w=\Delta N_{\rm in}/(\Delta A \Delta t)$ is given by 
\begin{align}
    J_w & =\frac{n_\chi}{(2\pi)^3}\int d^3p \, \Theta(-p_z-m_d)\, \Theta(-p_z) \, f(p,x)\left(-\frac{p_z}{|p|}\right) \nonumber \\
    & =\frac{n_\chi}{(2\pi)^3}\int_0^\pi p \, d\theta \int_0^{2\pi} p \sin{\theta} \, d\phi \int_0^\infty dp \, (-\cos{\theta})\Theta(-p \cos{\theta}-m_d)\Theta(-\cos{\theta}) f(p) \nonumber \\
    & =\frac{n_\chi}{(2\pi)^2}\int_0^{-1} d(\cos{\theta}) \, \cos{\theta} \int_{-\frac{m_d}{\cos{\theta}}}^\infty dp \, \frac{p^2}{e^{\tilde{\gamma}(\sqrt{p^2+(m_\chi^{\rm out})^2}-\tilde{v}p \cos{\theta})/T}+1} \nonumber \\
    & =\frac{n_\chi}{(2\pi)^2}\int_0^{-1}dx\, x \int_{-\frac{m_d}{x}}^\infty dp \frac{p^2}{e^{\tilde{\gamma}(\sqrt{p^2+(m_\chi^{\rm out})^2}-\tilde{v}p x)/T}+1}.
\end{align}
Now, the number of particles going inside the bubble with respect to the plasma frame is $n_{\rm in}=J_w/(\gamma_w v_w)$. Hence, the fraction of particles trapped outside the bubble can be written as 
\begin{equation}
    F(m^{\rm out}_\chi,m^{\rm in}_\chi,T,\tilde{v},v_w)=1-\frac{n_{\rm in}}{n_{\rm eq}}
    \label{Ftrap}
\end{equation}
\begin{figure}[ht]
    \centering
    \includegraphics[width=0.45\linewidth]{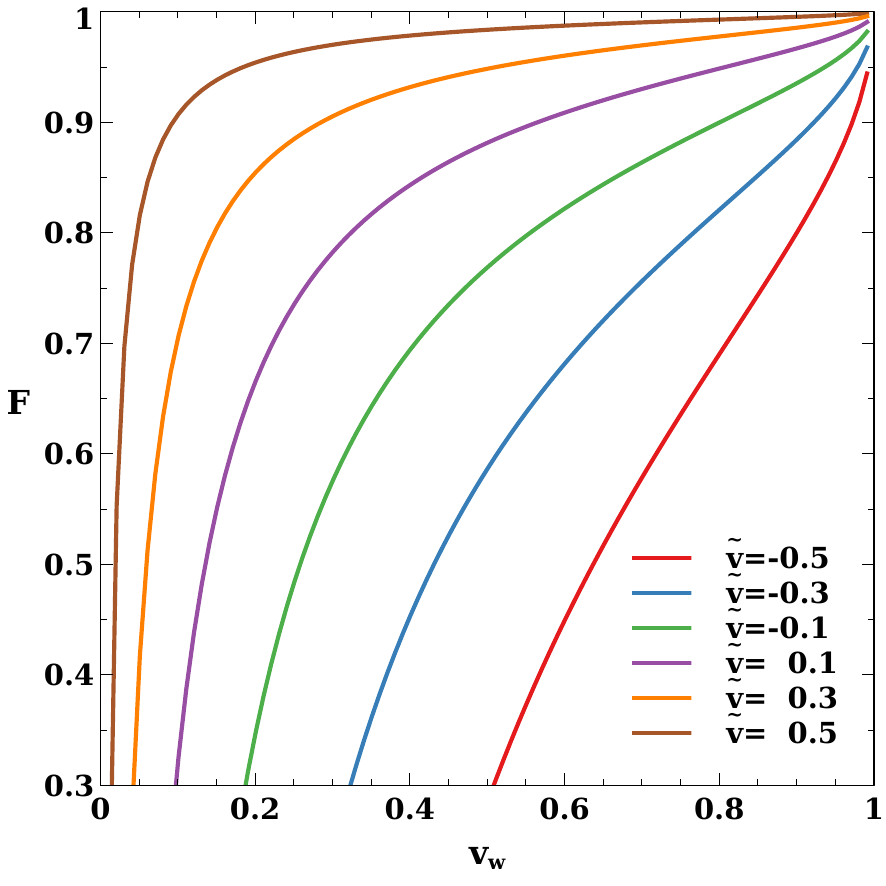}
     \includegraphics[width=0.45\linewidth]{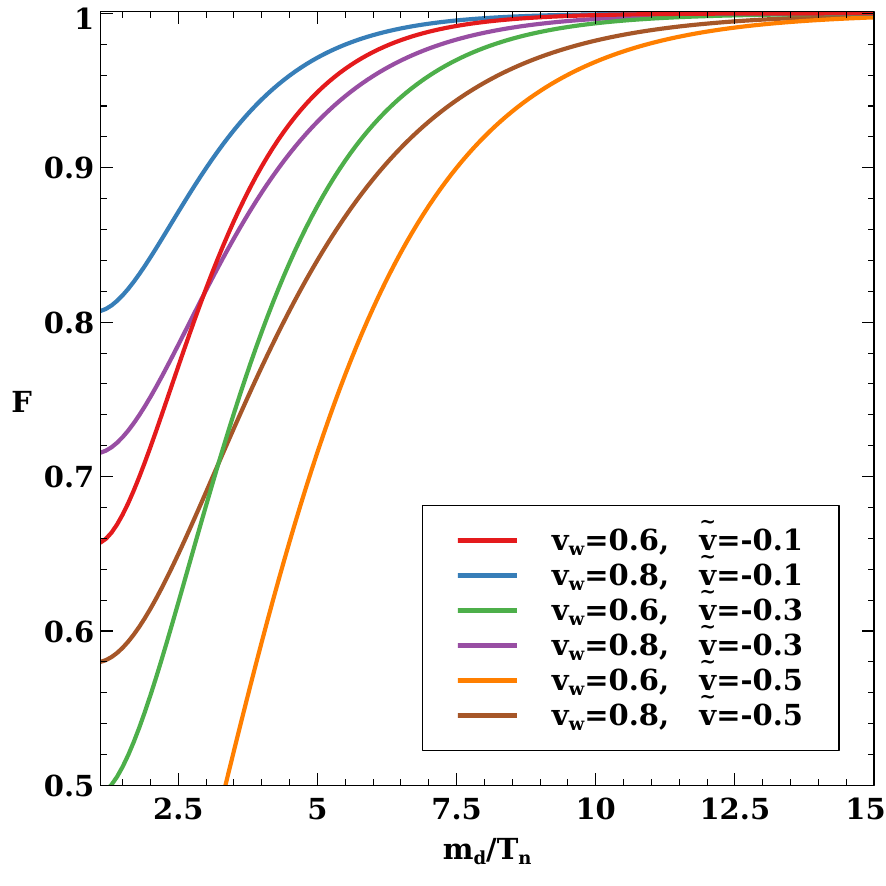}
    \caption{Left panel: Trapping fraction $F$ versus bubble wall velocity $v_w$ for different values of $\tilde{v}$, where $m^{\rm out}_\chi \sim T_n$ and $m_d/T_n\sim 2$. Right panel: Trapping fraction $F$ versus $m_d/T_n$ for different values of $\tilde{v}$ and $v_w$, where $m^{\rm out}_\chi\sim T_n$.}
    \label{fig1}
\end{figure}
From Fig.~\ref{fig1} left panel, it is evident that particle trapping is more efficient when the particles move away from the wall, corresponding to a positive $\tilde{v}$. Conversely, the trapping efficiency decreases for negative $\tilde{v}$. However, the trapping function increase with wall velocity, as seen from the right panel plot of Fig. \ref{fig1}. The right panel plot also shows the increase of the trapping fraction with the mass gap $m_d/T_n$ as expected. These also agree with the earlier works \cite{Chway:2019kft,Gehrman:2023qjn}. For a complete hydrodynamical treatment of bubble and fluid velocities, refer to appendix \ref{appen1}.

\section{Cogenesis of PBH and Baryon}
\label{sec3}
As the phase transition proceeds via bubble nucleation, the bubbles containing true vacuum grows to cover the entire Universe while the false vacuum regions shrink. Eventually, the false vacuum after shrinking to small disconnected volumes, split further to negligible size at percolation temperature \cite{Ellis:2018mja}. These squeezed pockets of false vacuum enhance the annihilation of dark fermions \cite{Asadi:2021pwo, Arakawa:2021wgz} leaving only the asymmetric part $\eta_\chi$. An approximate global charge at low scale ensures the survival of asymmetric dark fermion part in these false vacuum pockets leading to the degeneracy pressure. If the Fermi degeneracy pressure balances the vacuum pressure, it leads to the formation of Fermi-balls at $T_*$ defined by $p(T_*)=0.29$ \cite{Lee:1986tr, Hong:2020est,Kawana:2021tde,DelGrosso:2023trq}.

The energy of a Fermi-ball can be written as
\begin{equation}
    E_{\rm FB}=\frac{3\pi}{4}\left (\frac{3}{2\pi} \right)^{2/3}\frac{Q_{\rm FB}^{4/3}}{R} +\frac{4\pi}{3} U_0(T_*) R^3_{\rm FB},
\end{equation}
where, $Q_{\rm FB}=F_\chi^{\rm trap}\frac{n_\chi-n_{\Bar{\chi}}}{n_{\rm FB}^*}$ with $n_\chi$ denoting number density of $\chi$, $F_\chi^{\rm trap} \approx 1$ for maximum trapping of $\chi$ inside false vacuum and $U_0$ is vacuum energy. Now, minimizing the Fermi-ball energy, we can get the mass and radius as
\begin{align}
    M_{\rm FB}= Q_{\rm FB}(12\pi^2 U_0(T_*))^{1/4}, \quad \quad R_{\rm FB}^3=\frac{3}{16\pi}\frac{M_{\rm FB}}{U_0(T_*)}.
\end{align}
Within the Fermi-balls, the $\chi$'s  interact through attractive Yukawa interaction $g_\chi\phi\Bar{\chi}\chi$ and the interaction range is $L_\phi(T)= \left(\frac{d^2V_{\rm eff}}{d\phi^2}|_{\phi=0} \right)^{-1/2}$ \cite{Marfatia:2021hcp}. The Yukawa potential energy contribution to the Fermi-ball energy can be expressed as
\begin{equation}
    E_Y\simeq -\frac{3g_\chi^2}{8\pi}\frac{Q_{\rm FB}^2}{R_{\rm FB}} \left(\frac{L_\phi}{R_{\rm FB}}\right)^2.
\end{equation}
At a later stage, when the Yukawa potential energy becomes larger than the Fermi-ball energy, the Fermi-ball becomes unstable and collapses to form PBH \cite{Kawana:2021tde}. We consider that the Fermi-ball formation temperature is close to the nucleation temperature, $T_*\sim T_n $\footnote{We also find that the trapping function $F$ remains almost same at $T_n$ and $T_*$ justifying these temperatures to be used interchangeably.}. The initial PBH mass during formation can be estimated in terms of FOPT parameters as \cite{Kawana:2021tde}
\begin{equation}
    M_{\rm PBH}\sim 1.4\times10^{21} v_w^3 \left(\frac{\eta^{\rm trap}_\chi}{10^{-3}} \right) \left(\frac{100}{g_*}\right)^{1/4}  \left(\frac{100 \hspace{0.1 cm} \rm GeV}{T_n}\right)^{2}  \left(\frac{100}{\beta/\mathcal{H}_*}\right)^{3} \alpha_*^{1/4} \hspace{0.2 cm}{\rm g} \label{eq:mpbh}
\end{equation}
where $\eta_\chi^{\rm trap}=F\eta_\chi$ denotes the asymmetric part trapped inside the false vacuum pockets. Primordial black holes, if sufficiently heavy, can contribute to dark matter and its contribution is parametrised as \cite{Kawana:2021tde}
\begin{equation}
    f_{\rm PBH}=1.3*10^{3}v_w^{-3}\left(\frac{g_*}{100}\right)^{1/2}\left(\frac{T_n}{100\, \rm GeV}\right)^3\left(\frac{\beta/\mathcal{H}_*}{100}\right)^3\left(\frac{M_{\rm PBH}}{10^{15} \rm g}\right).
    \label{fdm}
\end{equation}
For PBHs constituting entire DM in the Universe, we have $f_{\rm PBH}=1$. PBHs can fully constitute dark matter within the mass range of $9.15\times10^{16}$ g and $6.70\times10^{21}$ g. The lower bound on mass comes from the evaporation constraint from the cosmic microwave background (CMB) observations \cite{Carr:2009jm,Lehmann:2018ejc} while the upper bound on mass comes from the micro-lensing constraint\cite{Niikura:2017zjd}.

In a co-genesis scenario, the initial dark sector asymmetry helps in the formation of PBH dark matter, while a portion of it filtered through the bubble walls is transferred to baryons or visible sector generating the observed baryon asymmetry of the Universe. As the bubble expands in the Universe, a fraction of the dark asymmetry remains outside the bubble, while the rest passes through the wall and subsequently decays into the visible sector. If the initial asymmetry is partitioned into two components, one contributing to the formation of PBHs, while the other fully converting into the baryon asymmetry, we can write $\eta_\chi =\eta_\chi^{\rm trap}+\eta_\chi^{\rm in}$ or $F+\frac{\eta_\chi^{\rm in}}{\eta_\chi}=1$.

\begin{figure}
    \centering
    \includegraphics[width=0.49\linewidth]{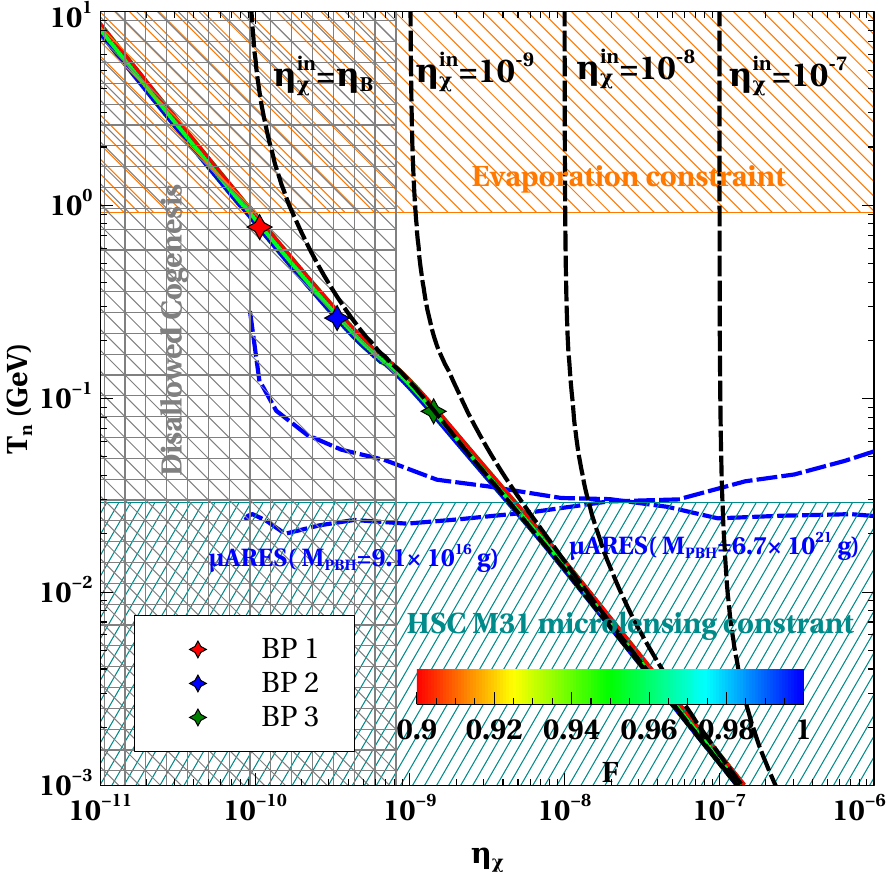}
        \includegraphics[width=0.49\linewidth]{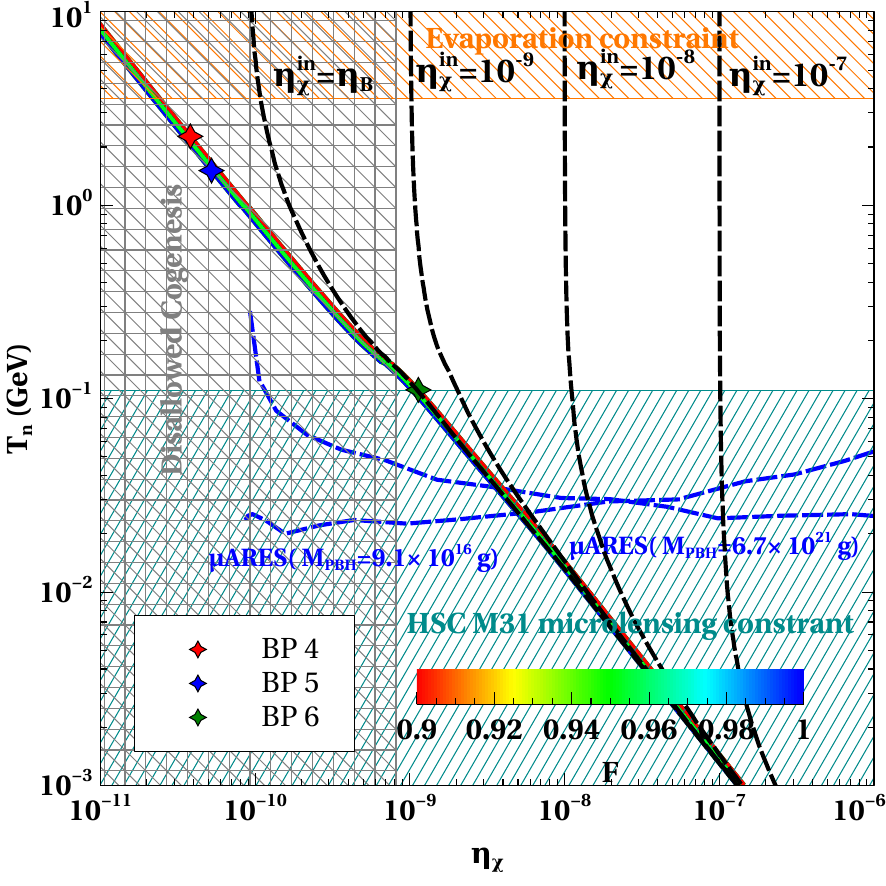}
    \caption{The parameter space in the $\eta_\chi$-$T_n$ plane where the solid green line represents condition where PBHs fully constitute dark matter ($f_{\rm PBH}=1$) consistent with appropriate bubble filtering. 
    The black dashed lines correspond to the parameter space with fixed dark asymmetry filtered to the true vacuum. For left panel, $\beta/\mathcal{H}_*=200$, $m_\chi^{\rm out} \sim T_n$, $\alpha_*=0.3$ and $v_w=0.85$ considering deflagration regime. Same for right panel with $\beta/\mathcal{H}_*=50$, $m_\chi^{\rm out} \sim T_n$, $\alpha_*=0.3$ and $v_w=0.85$.}
    \label{fig5}
\end{figure}

\begin{figure}[h!]
    \centering
    \includegraphics[width=0.49\linewidth]{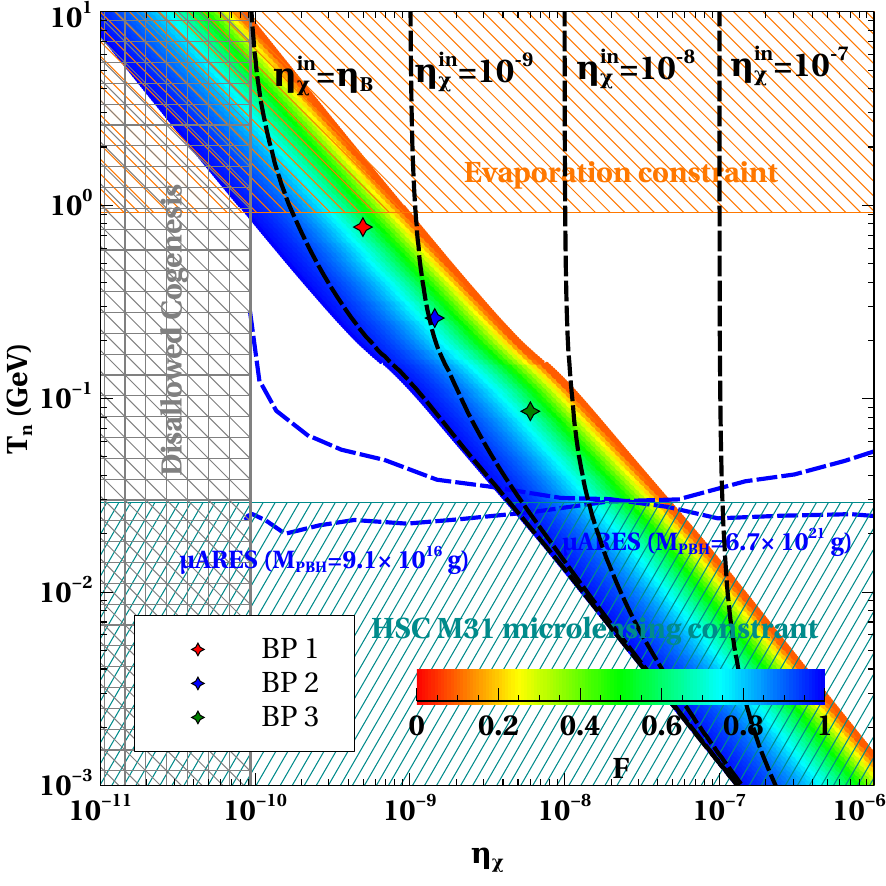}
        \includegraphics[width=0.49\linewidth]{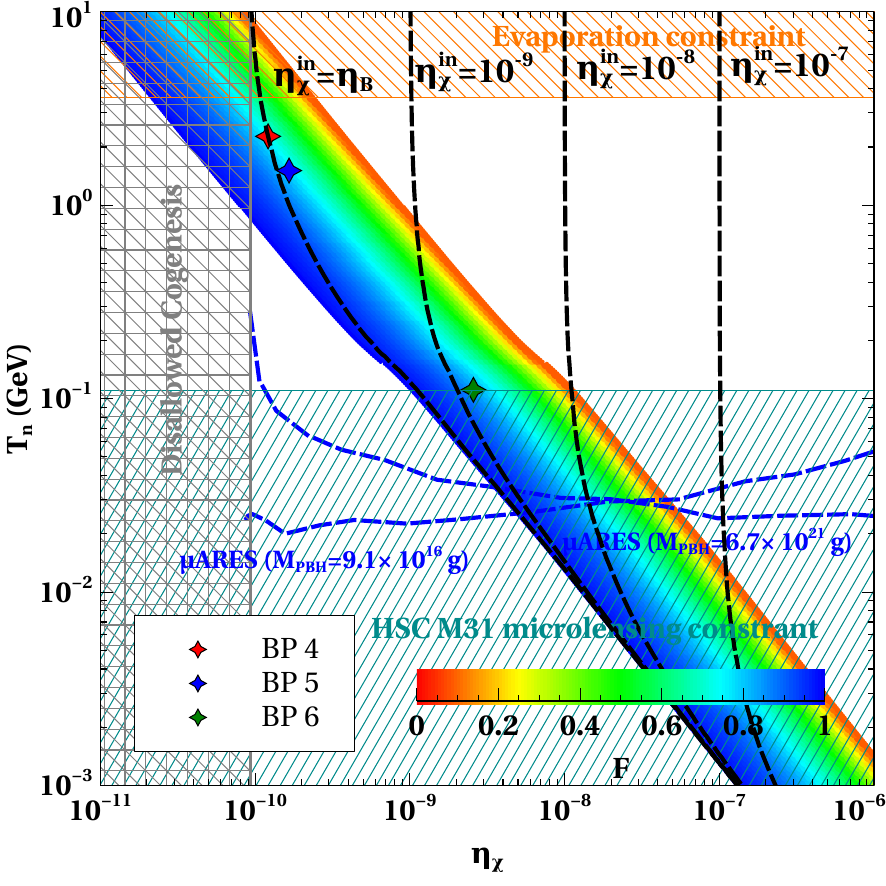}
    \caption{The parameter space in the $\eta_\chi$-$T_n$ plane where the solid green line represents condition where PBHs fully constitute dark matter ($f_{\rm PBH}=1$) consistent with appropriate bubble filtering. 
    The black dashed lines correspond to the parameter space with fixed dark asymmetry filtered to the true vacuum. For left panel, $\beta/\mathcal{H}_*=200$, $m_\chi^{\rm out} \sim T_n$, $\alpha_*=0.3$ and $v_w=0.85$ considering detonation regime. Same for right panel with $\beta/\mathcal{H}_*=50$, $m_\chi^{\rm out} \sim T_n$, $\alpha_*=0.3$ and $v_w=0.85$.}
    \label{fig6}
\end{figure}

In Fig.~\ref{fig5}, we present the parameter space in $T_n-\eta_\chi$ plane which satisfies the requirements of successful cogenesis of PBH dark matter and baryon asymmetry. We fix $\beta/\mathcal{H}_*=200$, $\alpha_*=0.3$, $v_w=0.85$ for the left panel and $\beta/\mathcal{H}_*=50$, $\alpha_*=0.3$, $v_w=0.85$ for the right panel of Fig.~\ref{fig5}. The colored band denotes the region where PBHs can account for entire DM. This can be found by using Eq. \eqref{eq:mpbh} in Eq. \eqref{fdm} to obtain
\begin{equation}
    f_{\rm PBH}=5.75 \times 10^8\alpha_*^{1/4}g^{1/4}_* \left (\frac{F\eta_\chi}{10^{-3}}\right )\left (\frac{T_n}{100} \right ) 
    \label{fpbh2}
\end{equation}
and then setting $f_{\rm PBH}=1$. We also calculate the trapping function $F \equiv F(m^{\rm out}_\chi,m^{\rm in}_\chi,T_n,\tilde{v},v_w)$ numerically using Eq. \eqref{Ftrap}, where $m_d/T_n$ is varied from 1 to 50 for each of these points in order to get one-to-one relation between $\eta_\chi$ and $T_n$ in deflagration regime. The black dashed contours are plotted by fixing dark fermion asymmetry leaked to the true vacuum via bubble filtering $\eta^{\rm in}_\chi$. This allows using $F\eta_\chi = \eta_\chi-\eta^{\rm in}_\chi$ in Eq. \eqref{fpbh2} leading to a relation between $\eta_\chi$ and $T_n$. This is clear from the asymptotic behavior of the black dashed contours as $\eta_\chi \rightarrow \eta^{\rm in}_\chi$. Only where black dashed contours meet the colored band, however, the trapping function $F$ corresponds to the exact value calculated in terms of the relevant model parameters $F(m^{\rm out}_\chi,m^{\rm in}_\chi,T_n,\tilde{v},v_w)$. To calculate the trapping fraction which is a crucial parameter for cogenesis, we consider the explicit dependence of $F$ on parameters related to the FOPT: $m^{\rm out}_\chi$, $m^{\rm in}_\chi$, $T_n$, and $v_w$. The remaining parameter is $\tilde{v}(v_f)$, the local fluid velocity outside the bubble wall in the wall frame, which is related to the fluid velocity inside the wall, $v_t$, through the energy-momentum conservation equations. We solve the hydrodynamical profile of the fluid velocity inside the wall for a given wall velocity in plasma frame, and then use a Lorentz transformation to relate $v_t$ with $v_w$. This allows us to determine $\tilde{v}$ as a function of $v_w$ the details of which is given in Appendix \ref{appen1}. Since the leftmost dashed contour corresponds to $\eta^{\rm in}_\chi = \eta_B$, the observed baryon asymmetry, the gray meshed region towards the left of the point where it touches the colored band is disfavored from cogenesis point of view. For this analysis, we find the local fluid velocity as a function of $v_w$, considering the deflagration regime as discussed in appendix \ref{appen1} and Fig. \ref{fig3} therein. In Fig.~\ref{fig6}, we show the parameter space in $T_n-\eta_\chi$ plane for the same set of parameters as in Fig.~\ref{fig5}, considering the detonation regime discussed in appendix \ref{appen1}. The detonation regime results in a high local fluid velocity relative to the wall, leading to a wider range of the trapping function $F$ ranging from $0.08$ to $1$. Consequently, the colored region, where PBHs constitute the entire DM, becomes broader compared to the deflagration regime of Fig. \ref{fig5}. 

The orange meshed region in Fig. \ref{fig5}, \ref{fig6} correspond to the disallowed PBH mass window on lighter side which affects big bang nucleosynthesis (BBN) and the extragalactic photon background \cite{Carr:2009jm,Lehmann:2018ejc}. Additionally, the light-green meshed region corresponding to heavier PBH mass is disfavored by HSC-M31 microlensing constraints \cite{Niikura:2017zjd}. All these constraints finally allow only the white colored regions in Fig. \ref{fig5}, \ref{fig6} with the cogenesis favored points lying on the green colored band. Different colored points on this band refer to the benchmark points given in table \ref{tab1} and table \ref{tab2}. The two blue dashed contours represent the $\mu$ARES sensitivity for PBH mass fixed at the minimum and maximum of the allowed range for it to be DM. Interestingly, all these benchmark points remain within the sensitivities of future GW experiments like $\mu$ARES \cite{Sesana:2019vho}. The stochastic GW spectrum from a FOPT is estimated by considering all the relevant contributions from bubble collisions \cite{Turner:1990rc,Kosowsky:1991ua,Kosowsky:1992rz,Kosowsky:1992vn,Turner:1992tz}, the sound wave \cite{Hindmarsh:2013xza,Giblin:2014qia,Hindmarsh:2015qta,Hindmarsh:2017gnf} and the turbulence \cite{Kamionkowski:1993fg,Kosowsky:2001xp,Caprini:2006jb,Gogoberidze:2007an,Caprini:2009yp,Niksa:2018ofa} of the plasma medium.

\begin{figure}[h!]
    \centering
    \includegraphics[width=0.49\linewidth]{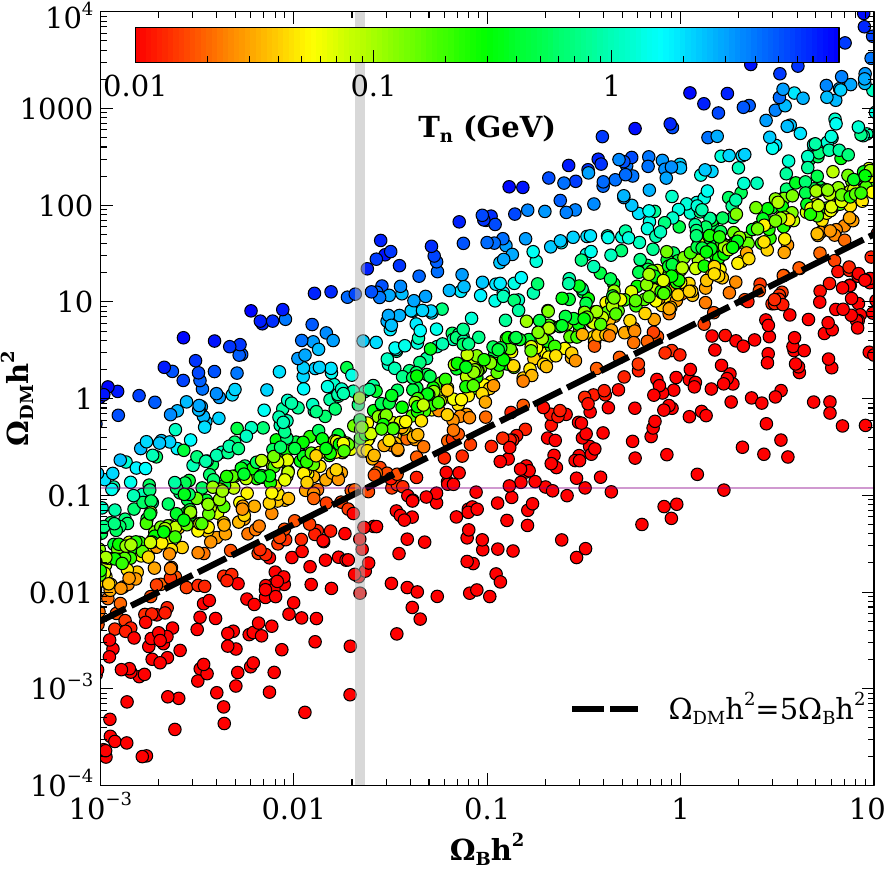}
        \includegraphics[width=0.49\linewidth]{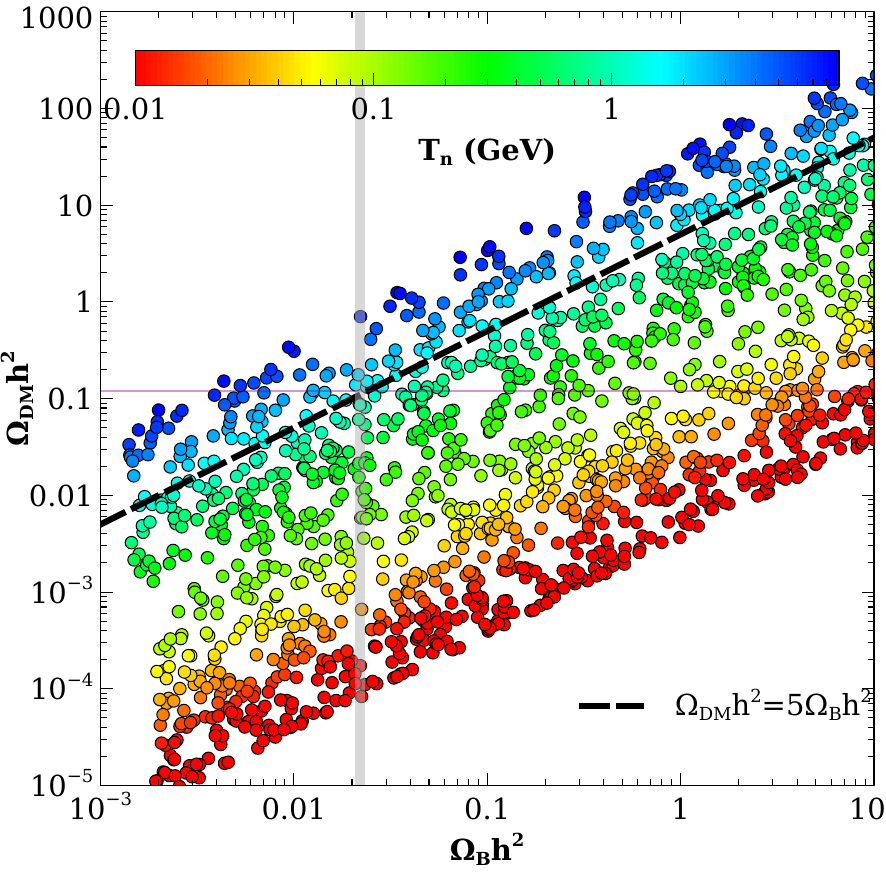}
    \caption{Variation of DM abundance $\Omega_{\rm DM}h^2$ with baryon abundance $\Omega_{\rm B}h^2$. The color bar indicates the nucleation temperature in the deflagration (left) and detonation (right) regimes. The horizontal and vertical bands correspond to the observational limits on DM and baryon densities respectively.}
    \label{fig6a}
\end{figure}
\begin{figure}
    \centering
    \includegraphics[width=0.5\linewidth]{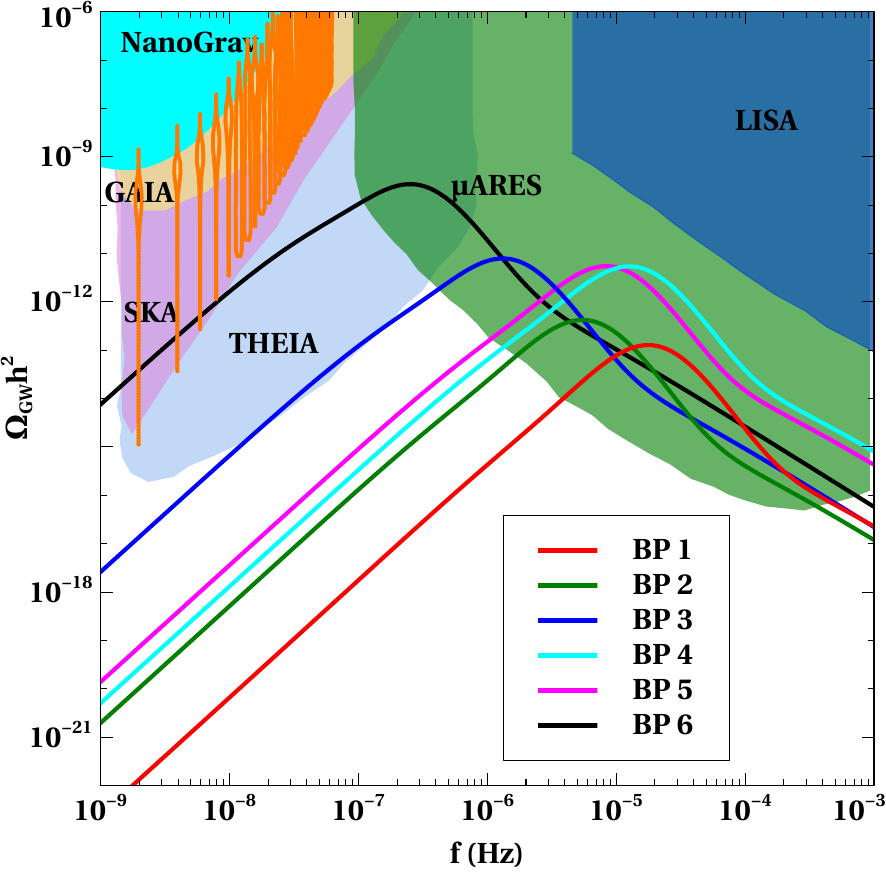}
    \caption{Gravitational wave spectrum corresponding to benchmark points in Table \ref{tab1} and \ref{tab2}.}
    \label{fig7}
\end{figure}

Clearly, $F$ varies over a wider range in the detonation regime. In the deflagration scenario, the benchmark points BP3 and BP6 satisfy both baryon asymmetry and PBH dark matter scenario. However, in detonation regime, BP4 satisfy both baryon asymmetry and PBH dark matter scenario, for other BPs baryon asymmetry is overproduced requiring additional washout in the true vacuum. For completeness, we perform a random scan satisfying the FOPT condition by varying $v_D$ from 10 MeV to 50 GeV, $m_\chi \leq 0.1v_D$ and $10^{-11} < \eta_\chi <10^{-6}$, while keeping $y_1 = 1.4$ and $\lambda_{\Phi_1\Phi_2} = 3.5$. The resulting points, correlating visible matter and dark matter abundances, are shown in left and right panels of Fig.~\ref{fig6a} for the deflagration and detonation regimes respectively. Since detonation regime can have wider allowed values of $F \leq 1$, it leads to smaller DM abundances compared to the deflagration. The black dashed line corresponds to DM abundance five times of baryons $\Omega_{\rm DM} =5 \Omega_B$, suggested by observations. Thus we observe a proportionate rise in DM abundance with baryon density while also accommodating the cogenesis motivated parameter space. Fig. \ref{fig7} shows the GW spectra for the chosen benchmark points given in table \ref{tab1} and \ref{tab2}. The sensitivities of planned future experiment like LISA \cite{2017arXiv170200786A}, $\mu$ARES \cite{Sesana:2019vho}, THEIA~\cite{Garcia-Bellido:2021zgu}, GAIA \cite{Garcia-Bellido:2021zgu}, SKA~\cite{Weltman:2018zrl} are shown as shaded regions together with the recent NANOGrav data~\cite{NANOGrav:2023gor} as orange colored violin-shaped points. Due to the restrictions of having the FOPT below the electroweak scale, all the benchmark points remain within the sensitivities of future experiments like $\mu$ARES \cite{Sesana:2019vho}.

\begin{table}[]
    \centering
    \resizebox{\columnwidth}{!}{
    \begin{tabular}{|c|c|c|c|c|c|c|c|c|c|c|}
    \hline
   &  $v_D$(GeV) & $m_{\Phi_2}$(GeV) & $m_\chi$(GeV) & $m_\chi^{\rm in}$(GeV) & $T_c$(GeV) & $T_n$(GeV) & $\beta/\mathcal{H}_*$ & $\alpha_*$ & $M_{\rm PBH}$ (g)\\
    \hline
    BP1 & 3.5   & 3.61 & 0.50 & 3.46 &   1.12 & 0.77 & 200 & 0.05 & $9.9\times10^{16}$\\
    \hline 
    BP2 & 1.2   & 1.24 & 0.20 & 1.21 &   0.38 & 0.26 & 200 & 0.08 & $3.3\times10^{18}$\\
    \hline
    BP3 & 0.4   & 0.42 & 0.06 & 0.39 &   0.12 & 0.08 & 200 & 0.26 & $2.7\times10^{20}$\\
    \hline
          
    \end{tabular} }
    \caption{Benchmark points for first-order phase transition.}
    \label{tab1}
\end{table}

\begin{table}[]
    \centering
    \resizebox{\columnwidth}{!}{
    \begin{tabular}{|c|c|c|c|c|c|c|c|c|c|c|}
    \hline
   &  $v_D$(GeV) & $m_{\Phi_2}$(GeV) & $m_\chi$(GeV) & $m_\chi^{\rm in}$(GeV) & $T_c$(GeV) & $T_n$(GeV) & $\beta/\mathcal{H}_*$ & $\alpha_*$ & $M_{\rm PBH}$ (g)\\
    \hline
    BP4 & 12 &   15.91 & 0.60 & 12.47 &  3.75 & 2.27 & 50 & 0.09 & $9.2\times10^{17}$\\
    \hline 
    BP5 & 8 &   10.62 & 0.40 & 8.31 &   2.50 & 1.51 & 50 & 0.09 & $2.9\times10^{18}$\\
    \hline
    BP6 & 0.6 &   0.79 & 0.10 & 0.69 &   0.18 & 0.11 & 50 & 0.42 & $1.8\times10^{22}$\\
    \hline
          
    \end{tabular} }
    \caption{Benchmark points for first-order phase transition.}
    \label{tab2}
\end{table}

\begin{figure}
    \centering
    \includegraphics[width=0.45\linewidth]{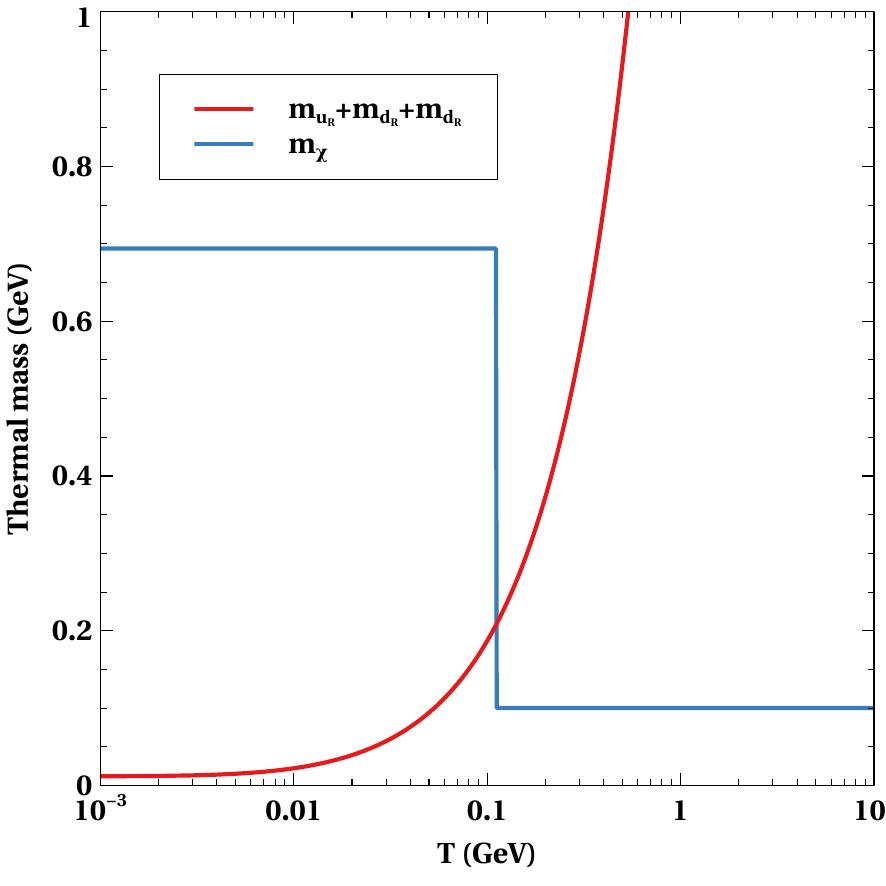}
        \includegraphics[width=0.5\linewidth]{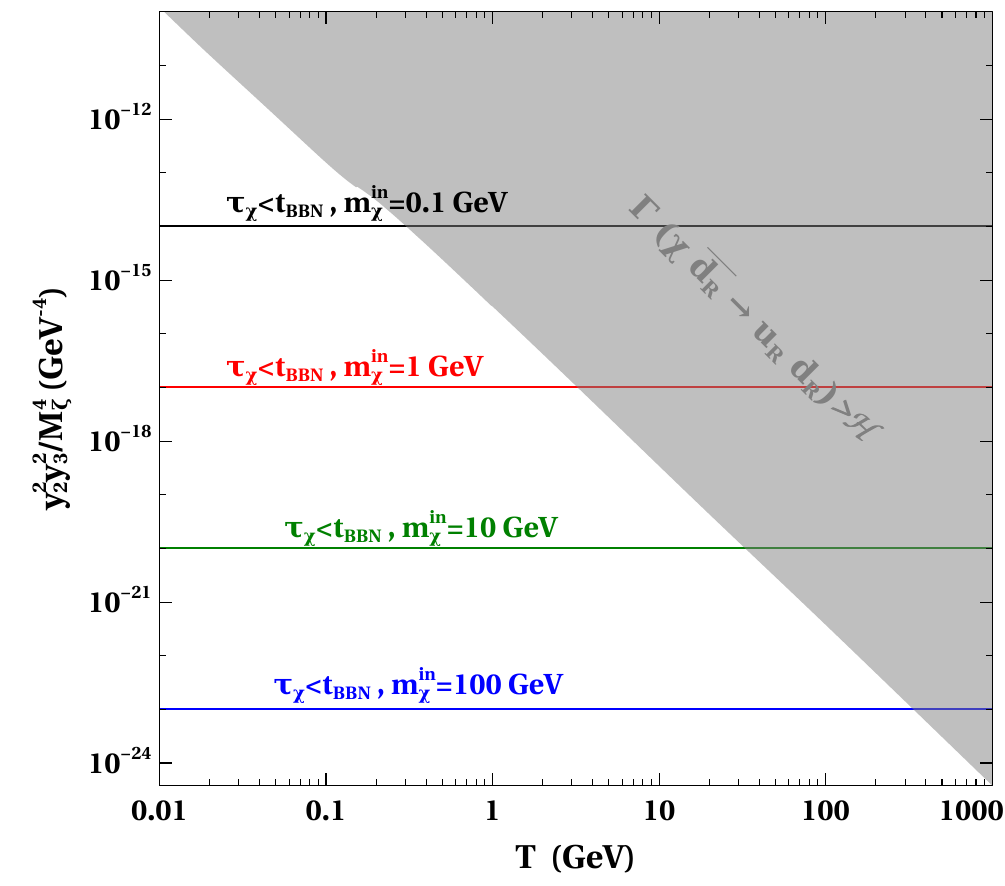}
    \caption{Left panel: Thermal mass profiles of dark fermion $\chi$ and quarks corresponding to the benchmark point BP6 in Table \ref{tab2}. Right panel: Allowed parameter space from the criteria of lifetime $\tau_\chi < t_{\rm BBN}$ and inefficient transfer of dark fermion asymmetry to baryons via scattering. The white region above a horizontal line is allowed for a specific value of $\chi$ mass in true vacuum.}
    \label{fig8}
\end{figure}

If the phase transition and bubble filtering of $\chi$ occurs before sphaleron decoupling, $\chi$ can transfer the asymmetry into leptons which can then get converted into baryon asymmetry via electroweak sphalerons. However, as shown in Fig. \ref{fig5} and Fig. \ref{fig6}, the allowed region consistent with PBH DM corresponds to nucleation temperature only upto a few GeV. This requires the dark fermion asymmetry to be transferred directly into baryons in the true vacuum. This is possible by identifying $\psi$ in Eq. \eqref{eq:yukawa} as a right-handed down-type quark $d_R$ with $\zeta$ being a colored scalar with electromagnetic charge $-1/3$. This lead to the following interactions relevant for transferring dark fermion asymmetry:
\begin{equation}
    -\mathcal{L} \supset y_2 \overline{d_R} \zeta \chi + y_3 \zeta^* \overline{u^c_R} d_R + {\rm h.c.}.
\end{equation}
Similar interactions have been considered in low scale baryogenesis models \cite{Aitken:2017wie}. While $\zeta$ is much heavier than $\chi$ or SM particles due to collider bounds on such a colored scalar \cite{Aitken:2017wie}, we consider three body decay $\chi \rightarrow u_R d_R d_R $ mediated by off-shell $\zeta$ to transfer the dark fermion asymmetry to the baryons in the true vacuum.

While $\chi$ is stable in the false vacuum prior to FOPT or the formation of PBH, it can decay into three SM quarks due to the sudden change in mass after the phase transition. In the left panel of Fig. \ref{fig8}, we show the variation of finite-temperature masses of $\chi$ and combination of three right-handed quarks. The thermal mass of quarks can be written as \cite{Bellac:2011kqa,Laine:2016hma}
\begin{equation}
    m_{u_R,d_R}^2(T)=m_{u_R,d_R}^2 (T=0)+\frac{1}{6}g_s^2T^2+ \frac{1}{8}4\pi\alpha_{\rm em}T^2.
\end{equation}
Clearly, before the FOPT, $\chi$ decay is kinematically forbidden while in a true vacuum, $\chi$ can decay such that it transfers the remnant asymmetry to baryons. The corresponding three-body decay width is given by
\begin{equation}
    \Gamma_\chi \equiv \Gamma (\chi \rightarrow u_R d_R d_R) \approx  \frac{ y^2_2 y^2_3(m^{\rm in}_\chi)^5}{2038 \pi^3 M^4_\zeta},
    \label{yDbound}
\end{equation}
where $M_\zeta$ denotes the mass of the colored scalar $\zeta$. Since dark fermion stores an asymmetry much larger than the observed baryon asymmetry, it is also required to ensure that they do not transfer the asymmetry to quarks via scattering
processes like $\chi \overline{d_R} \rightarrow u_R d_R$ mediated by $\zeta$. Keeping this conversion process out-of-equilibrium at a temperature $T$ leads to
\begin{equation}
T^5 \frac{y^2_2 y_3^{2}}{64 \pi M^4_\zeta} \lesssim \sqrt{\frac{\pi^2}{90}g_*} \frac{T^{2}}{M_{\rm Pl}},\label{eq:washout2}
\end{equation}
assuming massless initial and final states appropriate for high temperatures. The out-of-equilibrium condition of this process can be satisfied for suitable choices of Yukawa couplings $y_2, y_3$ and colored scalar mass $M_\zeta$. However, the combination of this out-of-equilibrium criterion and lifetime of $\chi$: $\tau_\chi = \Gamma^{-1}_\chi < t_{\rm BBN}$ being shorter compared to the BBN timescale restrict the reheat temperature of the Universe (or the production temperature of dark fermion asymmetry $T_{\Delta \chi} < T_{\rm RH}$) to be $\lesssim \mathcal{O}(10)$ GeV depending upon the mass of $\chi$ in true vacuum. This requires the creation of dark fermion asymmetry just before the FOPT without leaving much window to transfer the asymmetry to quarks via scatterings. In the right panel of Fig. \ref{fig8}, we show $\frac{y^2_2 y_3^{2}}{M^4_\zeta}$ as a function of $T$ indicating the disfavored region (gray-shaded) where dark fermion asymmetry can be transferred to baryons via scattering. The region above the horizontal line for a particular value of $\chi$ mass in true vacuum corresponds to required $\chi$ lifetime being shorter than the BBN epoch $\tau_\chi < t_{\rm BBN}$ for successful baryogenesis. Clearly, depending upon $\chi$ mass in true vacuum, the parameter space of the model and the production temperature of dark fermion asymmetry are tightly constrained.

\section{Conclusion}
\label{sec5}
We have proposed a novel cogenesis mechanism of producing dark matter and baryon asymmetry in the Universe by utilising a first-order phase transition in the early Universe. The FOPT driven by a singlet scalar also leads to a sharp rise in a dark fermion mass. While the dark fermion remains in equilibrium prior to the FOPT, a large mass gap across the bubble wall $m_d \gg T_n$ prevents most of these dark fermions to get trapped in the false vacuum. An initial dark sector asymmetry prevents the dark fermions from being annihilated away in the squeezed pockets of false vacuum. While such false vacuum pockets lead to the formation of Fermi-balls when Fermi degeneracy pressure balances the vacuum pressure, a strong attractive Yukawa type interaction among dark fermions can induce further collapse to primordial black holes. Since the dark fermion momenta follow equilibrium distribution prior to the FOPT, its high momentum modes are energetic enough to leak into the true vacuum through the bubble walls, leading to filtering of dark fermions. The filtered dark fermion can also decay into the visible sector due to the mass gain in the true vacuum, thereby transferring the filtered dark sector asymmetry to visible sector asymmetry. We constrain the model from the requirement of producing the observed dark matter relic in the form of PBHs and observed baryon asymmetry from filtered dark sector asymmetry in the true vacuum. We find a small allowed region of parameter space with nucleation temperature ranging from a few tens of MeV to a few GeV which is consistent with the cogenesis requirements and experimental bounds on PBH mass. 
Interestingly, for this allowed region of parameter space, the FOPT generated stochastic gravitational wave remains within reach of future GW experiments like $\mu$ARES \cite{Sesana:2019vho}. Additionally, such low scale FOPT scenario has several light new degrees of freedom accessible at terrestrial experiments. The PBHs in this asteroid mass range can also have their own observational signatures related superradiance, Hawking evaporation or microlensing \cite{Dent:2024yje}. While we have performed our detailed numerical analysis assuming an initial dark sector asymmetry, we also outline one possible UV completion where such an asymmetry can be generated at a scale above the phase transition scale via the Affleck-Dine mechanism. If the FOPT occurs prior to the sphaleron decoupling epoch, the dark sector asymmetry can be transferred to leptons opening up the leptogenesis route. However, this will overproduce PBH in the desired mass range requiring additional entropy dilution. Alternately, we can have such high scale FOPT and production of heavier planetary mass PBH with $f_{\rm PBH} \ll 1$ which can show up in gravitational microlensing observations \cite{Mroz:2024wia}. While this allows PBH to be a small fraction of DM, the remaining DM can originate from leftover dark fermion asymmetry after partial transfer to baryon asymmetry. This requires different model building approach which we leave for future works.

\acknowledgements

We thank Arnab Dasgupta for useful discussions. The work of D.B. is supported by the Science and Engineering Research Board (SERB), Government of India grants MTR/2022/000575 and CRG/2022/000603. D.B. also acknowledges the support from the Fulbright-Nehru Academic and Professional Excellence Award 2024-25. 

\appendix 
\section{Hydrodynamical treatment of velocities}
\label{appen1}
As the first-order phase transition progresses, the expanding bubble releases latent heat into the radiation bath of the Universe, generating velocity and temperature profiles of the fluid across the bubble wall. In this section, we discuss the dynamics of wall velocity. The energy-momentum tensor of the scalar field that drives the phase transition is
\begin{equation}
T_{\alpha\beta}^\phi=\partial_\alpha\phi\partial_\beta\phi-g_{\alpha\beta}\left[ \frac{1}{2} \partial_\mu\phi\partial^\mu\phi-V_{\rm eff}(\phi)\right].
\end{equation}
On the other hand, the energy-momentum tensor of the plasma is given by
\begin{equation}
    T_{\alpha\beta}^{\rm pl}=\sum_i\int \frac{d^3k}{(2\pi)^3E_i}k_\alpha k_\beta f_i(k,x),
\end{equation}
where, the i index corresponds to each species in the plasma. Assuming the plasma to be locally in thermal equilibrium, its energy-momentum tensor can be expressed as that of a perfect fluid,
\begin{equation}
    T_{\alpha\beta}^{\rm pl}=w_{\rm pl}u_\alpha u_\beta-g_{\alpha\beta}p_{\rm pl},
\end{equation}
where $w_{\rm pl}$, $p_{\rm pl}$ and $u_\alpha \equiv \gamma(1,\textbf{\rm v})$ are plasma enthalpy, pressure and four-velocity respectively. With the $\phi$ background contributing to total pressure, the total fluid energy-momentum tensor can be written as
\begin{equation}
    T_{\alpha\beta}^{\rm fl}=T_{\alpha\beta}^\phi+T_{\alpha\beta}^{\rm pl}=w u_\alpha u_\beta-g_{\alpha\beta}p,
\end{equation}
where, the enthalpy $w=T\frac{\partial p}{\partial T}$ and the energy density $e=T\frac{\partial p}{\partial T}-p=w-p$. The conservation of energy-momentum leads to $\partial^\alpha T_{\alpha\beta}^{\rm fl}=0$. For a constant bubble wall velocity, the time-independent components of the energy-momentum conservation equations in bubble wall frame can be expressed as \cite{Espinosa:2010hh,Jiang:2023nkj}
\begin{equation}
    \partial_zT^{zz}=\partial_zT^{z0}=0
\end{equation}
where $z$-axis is chosen to be direction of bubble and fluid velocities. From the above equation, we get the matching equations across the bubble wall as follows
\begin{align}
    w_f v_f^2\gamma_f+p_f=w_t v_t^2\gamma_t+p_t \quad {\rm and} \quad w_f v_f \gamma_f^2=w_t v_t \gamma_t^2, \label{matching}
\end{align}
here, subscripts $t$ and $f$ represent inside the bubble (true vacuum) and outside the bubble (false vacuum), respectively. In the relativistic gas approximation of plasma, the equation of state outside the bubble can be written as
\begin{align}
    p_f=\frac{1}{3}a_fT_f^4-\epsilon, \quad e_f=a_fT_f^4+\epsilon,
    \label{bag1}
\end{align}
where, $\epsilon$ is the false vacuum energy and $a$ correspond to degrees of freedom across the wall. While inside the bubble
\begin{align}
    p_t=\frac{1}{3}a_tT_t^4, \quad e_t=a_tT_t^4. \label{bag2}
\end{align}
\begin{figure}
    \centering
    \includegraphics[width=0.5\linewidth]{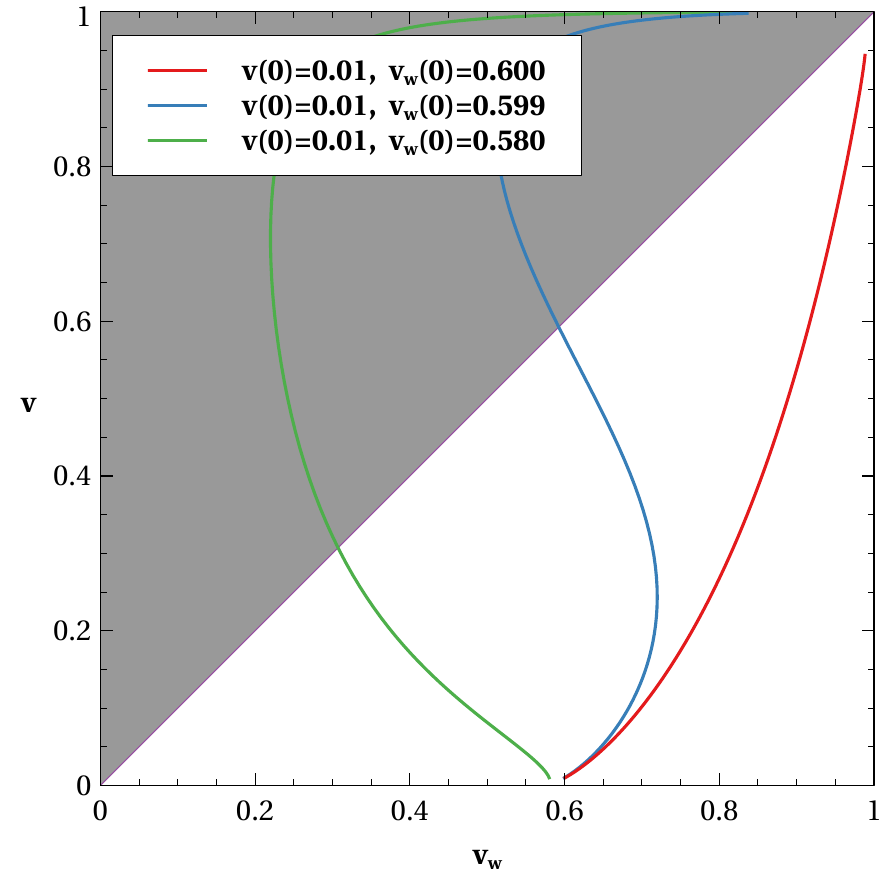}
    \caption{Generic fluid velocity profile with corresponding wall velocity $v_w=\xi$.}
    \label{fig2}
\end{figure}
\begin{figure}
    \centering
    \includegraphics[width=0.45\linewidth]{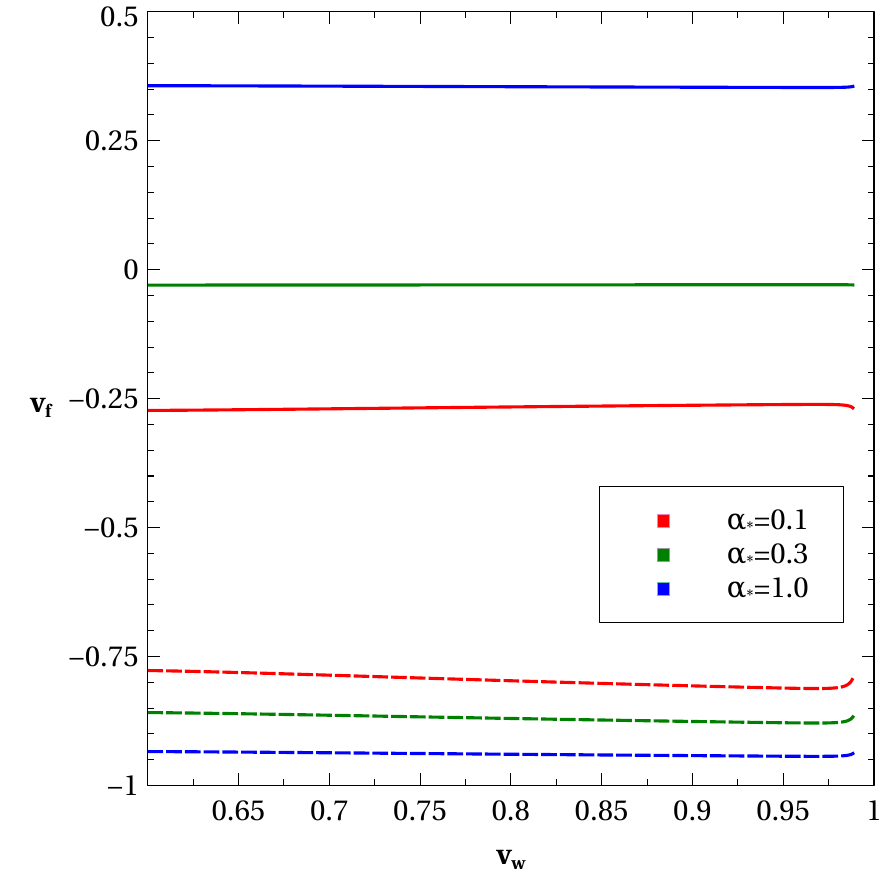}
    \includegraphics[width=0.45\linewidth]{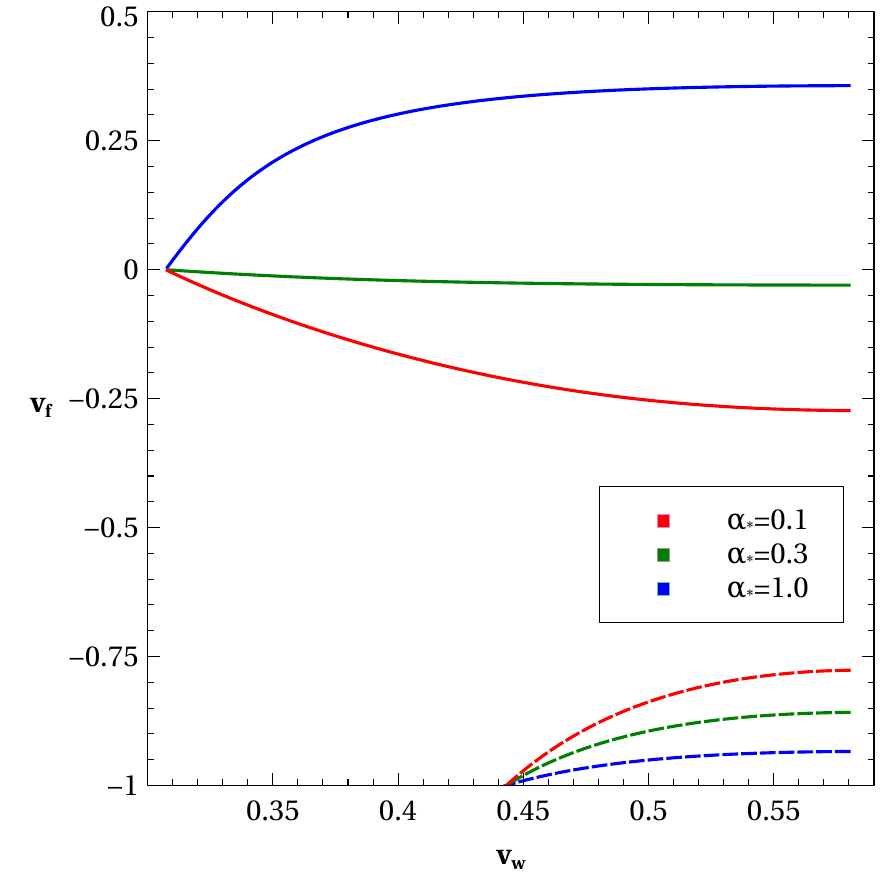}
    \caption{Left panel: local fluid velocity in front of the wall, $v_f$ versus bubble wall velocity for different values of $\alpha_*$ corresponding to the red colored profile of Fig. \ref{fig2}. Right panel: same as the left panel but corresponding to the green colored profile of Fig. \ref{fig2}. The solid and dashed lines correspond to deflagration and detonation regimes, respectively.}
    \label{fig3}
\end{figure}
After substituting Eq. \eqref{bag1} and Eq. \eqref{bag2} in Eq. \eqref{matching}, we get
\begin{align}
    v_fv_t =\frac{1-(1-\alpha_*)\rho}{3-3(1+\alpha_*)\rho}, \,\,\,\,
    \frac{v_f}{v_t}=\frac{3+(1-3\alpha_*)\rho}{1+3(1+\alpha_*)\rho},
\end{align}
where, $\alpha_*=\epsilon/(a_fT_f^4)$, $\rho=a_fT_f^4/(a_tT_t^4)$. Then, combining the above two equations, we get
\begin{equation}
    v_f=\frac{1}{1+\alpha_*}\left[ \left(\frac{v_t}{2}+\frac{1}{6v_t}\right)\pm \sqrt{\left(\frac{v_t}{2}+\frac{1}{6v_t}\right)^2+\alpha_*^2+\frac{2}{3}\alpha_*-\frac{1}{3}} \right]. \label{local_v}
\end{equation}
 There are two branches of solutions corresponding to the $\pm$ signs on the right hand side of Eq.~\eqref{local_v}. The positive branch solution belongs to detonation, $v_f>v_t$ and the negative branch solution gives deflagration, $v_f<v_t$. Connecting with the discussion in section \ref{sec2}, $v_f$ represents the velocity of the fluid in front of the wall i.e. $\tilde{v}$, while $v_t$ denotes the velocity of the fluid behind the wall, both measured in the wall rest frame. We now calculate the bubble wall velocity and the corresponding velocity of the fluid moving behind the wall from plasma frame of reference. The conservation equation can be written as
\begin{equation}
    \partial_\alpha T^{\alpha\beta}=u^\beta\partial_\alpha(u^\alpha w)+u^\alpha w\partial_\alpha u^\beta-\partial^\beta p.
\end{equation}
Taking the projection along the flow, leads to
\begin{equation}
    \partial(u^\alpha w)-u_\alpha \partial^\alpha p=0. \label{proj}
\end{equation}
Now, taking the projection perpendicular to the flow with $\Bar{u}_\alpha\equiv \gamma(v,\textbf{\rm v}/v)$, we get
\begin{equation}
    \Bar{u}^\alpha u^\beta w \partial_\alpha u_\beta-\Bar{u}^\beta\partial_\beta p=0. \label{perpen_proj}
\end{equation}
In the absence of a characteristic distance scale, the solution is expected to depend on the self-similar parameter $\xi=r/t$, where $r$ is the distance from the center of the bubble, and $t$ is the time elapsed since nucleation. The gradients can be defined as
\begin{align}
    u_\alpha\partial^\alpha=-\frac{\gamma}{t}(\xi-v)\partial_\xi, \quad  \Bar{u}_\alpha\partial^\alpha=\frac{\gamma}{t}(1-\xi v)\partial_\xi.
\end{align}
From Eq. \eqref{proj} and Eq. \eqref{perpen_proj}, we get
\begin{align}
    (\xi-v)\frac{\partial_\xi e}{w} =2\frac{v}{\xi}+\{1-\gamma^2v(\xi-v)\}\partial_\xi v, \,\,\,\,
    (1-v\xi)\frac{\partial_\xi p}{w} =\gamma^2(\xi-v)\partial_\xi v.
\end{align}

\begin{figure}
    \centering
    \includegraphics[width=0.45\linewidth]{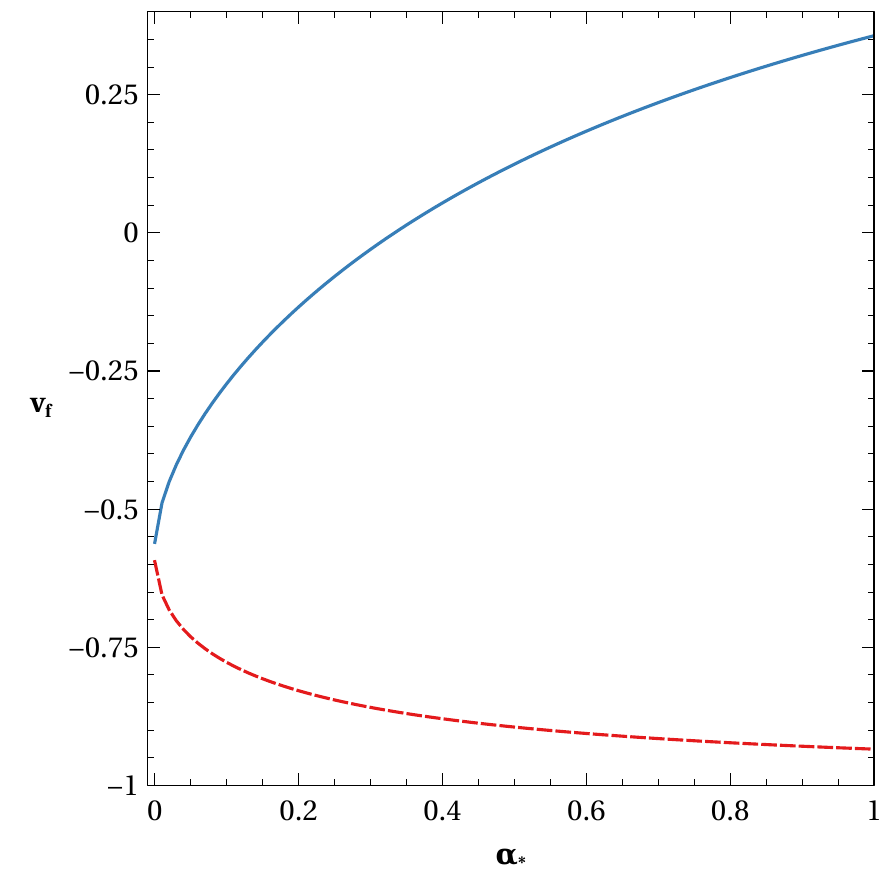}
    \includegraphics[width=0.45\linewidth]{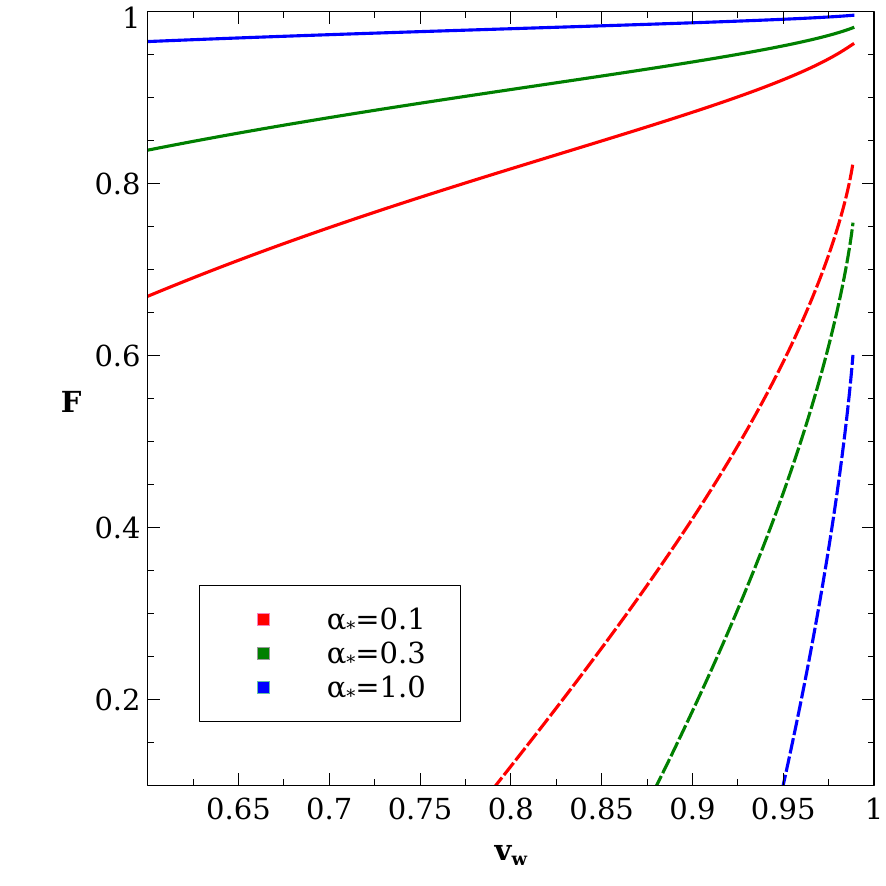}
    \caption{Left panel: Local fluid velocity in front of wall, $v_f$ versus $\alpha_*$, Right panel: Fraction of trapped particles $F$ versus bubble wall velocity, $v_w$ corresponding to the left panel of Fig.~\ref{fig3}. The solid and dashed lines correspond to deflagration and detonation regimes, respectively.}
    \label{fig4}
\end{figure}

Using the definition of the sound speed of the plasma $c_s^2\equiv(dp/dT)/(de/dT)$, the above two equations can be expressed as
\begin{equation}
    2\frac{v}{\xi}=\gamma^2(1-v\xi)(\frac{\mu^2}{c_s^2}-1)\partial_\xi v,
\end{equation}
where, $\mu=(\xi-v)/(1-\xi v)$. The above equation can be solved by introducing an auxiliary parameter $\tau$, which helps to write the equation as
\begin{align}
    \frac{dv}{d\tau} =2vc_s^2(1-v^2)(1-\xi v), \,\,\,\,
    \frac{d\xi}{d\tau} =\xi((\xi-v)^2-c_s^2(1-\xi v)^2),
\end{align}
and the solutions for various initial conditions are illustrated in Fig.~\ref{fig2}. Here, we show three types of profiles corresponding to the green, blue and red colored contours respectively. The gray shaded upper triangular region in the does not provide physically consistent solutions.

The fluid velocity behind the wall is $v$ in the plasma frame, whereas it is $v_t$ in the wall's rest frame. We can write $v=(v_t+v_w)/(1+v_tv_w)$ using Lorentz transformation of velocities. Thus, $v_w$ can be expressed as a function $f(v_f)$ in terms of the local plasma velocity in front of the wall, measured in the wall's rest frame. In Fig. \ref{fig3}, we connect the local fluid velocity with wall velocity, with solid and dashed lines corresponding to deflagration and detonation regimes, respectively. From the left panel of Fig.~\ref{fig3}, we can see that $v_f$ is nearly independent of the wall velocity $v_w$, while it depends on $\alpha_*$ shown in the left panel of Fig.~\ref{fig4}. In the right panel of Fig.~\ref{fig4}, we show the fraction of particles trapped outside the bubble as a function of wall velocity, where $m_\chi^{\rm out} \sim T_n$, $m_\chi^{\rm in} \sim 2T_n$ corresponding to the left panel of Fig.~\ref{fig3}. We use these dependence of fluid and bubble velocities while calculating the trapping function $F$ relevant for cogenesis discussed in section \ref{sec3}.

\bibliographystyle{JHEP}
\bibliography{ref1,ref2} 

\end{document}